\newcommand{\be}{\begin{equation}}
\newcommand{\ee}{\end{equation}}
\newcommand{\bea}{\begin{eqnarray}}
\newcommand{\eea}{\end{eqnarray}}

\newcommand{\fig}[1]{Fig.~(\ref{#1})}
\newcommand{\eq}[1]{Eq.~(\ref{#1})}

\newcommand{\kB}{k_{\rm B}}
\newcommand{\kT}{k_{\rm B}T}

\newcommand{\Hamil}{{\cal H}}
\newcommand{\Z}{{\cal Z}}

\newcommand{\la}{\left\langle}
\newcommand{\ra}{\right\rangle}

\newcommand{\tw}{t_{\rm w}}

\newcommand{\Ed}{E_{\rm d}}
\newcommand{\Ea}{E_{\rm a}}
\newcommand{\wij}{\omega_{ij}}
\newcommand{\wkl}{\omega_{kl}}
\newcommand{\fij}{f_{ij}}
\newcommand{\G}{G}
\newcommand{\xid}{\xi_{\rm d}}
\newcommand{\hd}{h_{\rm d}}
\newcommand{\Ms}{M_{\rm s}}
\newcommand{\D}{A}
\newcommand{\obs}{B}
\newcommand{\dW}{d\Gamma}
\newcommand{\raa}{\right\rangle_{\rm a}}
\newcommand{\Wa}{W_{\rm a}}
\newcommand{\Za}{Z_{\rm a}}
\newcommand{\qikjl}{q_{ik:jl}}
\newcommand{\qiljk}{q_{il:jk}}
\newcommand{\Pa}{\frac{1-S_{2}}{3}}
\newcommand{\Beff}{\vec{H}_{\rm eff}}
\newcommand{\bfl}{\vec{b}_{\rm fl}}
\newcommand{\tD}{\tau_{{\rm D}}}

\newcommand{\Qls}{{\cal C}}
\newcommand{\Rls}{\bar{\cal R}}
\newcommand{\Rns}{{\cal R}}
\newcommand{\Sls}{{\cal S}}
\newcommand{\Tls}{{\cal T}}
\newcommand{\Uls}{{\cal U}}
\newcommand{\Vls}{(\bar{\cal R}-{\cal R})}
\newcommand{\coeff}{a}

\newcommand{\Gij}{{\bf G}_{ij}}

\newcommand{\Gjk}{{\bf G}_{jk}}
\newcommand{\I}{{\bf 1}}

\newcommand{\rij}{\vec{r}_{ij}}
\newcommand{\vij}{\hat{r}_{ij}}
\newcommand{\m}{\vec{m}}
\newcommand{\n}{\vec{n}}

\newcommand{\e}{\vec{s}}
\newcommand{\nii}{\vec{n}_{i}}

\newcommand{\ei}{\vec{s}_{i}}
\newcommand{\ej}{\vec{s}_{j}}

\newcommand{\HH}{\vec{H}}

\newcommand{\h}{\hat{h}}
\newcommand{\z}{\hat{z}}

\newcommand{\Tg}{T_{\rm g}}
\newcommand{\chieq}{\chi_{\rm eq}}

\newcommand{\tobs}{t_{\rm obs}}

\newcommand{\Lovlp}{L_{\rm \Delta T}}

\newcommand{\case}[2]{{\textstyle \frac{#1}{#2}}}

\newcommand{\lesssim}
{\,\raisebox{0.35ex}{$<$}
\hspace{-1.7ex}\raisebox{-0.65ex}{$\sim$}\,
}

\documentclass[a4paper,11pt]{article}

\usepackage{citesort}

\usepackage{graphicx}
\usepackage{epsf}
\begin{document}


\title{\bf Superparamagnetism and Spin Glass Dynamics of
Interacting Magnetic Nanoparticle Systems}

\author{Petra E. J{\"o}nsson\\
\\
Department of Materials Science, Uppsala University, \\
Box 534, SE-751 21 Uppsala, Sweden \\
\\
Present address: Institute for Solid State Physics, University of Tokyo\\
Kashiwa-no-ha 5-1-5, Kashiwa, Chiba 277-8581, Japan \\
email: petra@issp.u-tokyo.ac.jp}
\maketitle

\newpage

\tableofcontents

\vspace{2cm}

\noindent
The physical properties of magnetic nanoparticles have been investigated with focus on the influence of dipolar interparticle interaction.
For weakly coupled nanoparticles, thermodynamic perturbation theory is employed to derive analytical expressions for the linear equilibrium susceptibility, the zero-field specific heat and averages of the local dipolar fields.
By introducing the averages of the dipolar fields in an expression for the relaxation rate of a single particle, a nontrivial dependence of the superparamagnetic blocking on the damping coefficient is evidenced. 
This damping  dependence is interpreted in terms of the nonaxially symmetric potential created by the transverse component of the dipolar field.

Strongly interacting nanoparticle systems are investigated experimentally in terms of spin glass behavior. 
Disorder and frustration arise in samples consisting of frozen ferrofluids from the randomness in particle position and anisotropy axis orientation.
A  strongly interacting FeC system is shown to exhibit critical dynamics characteristic of a spin glass phase transition.
Aging, memory and rejuvenation phenomena similar to those of conventional spin glasses are observed, albeit with much weaker rejuvenation  effects than in both a
 \mbox{Ag(11 at\% Mn)} Heisenberg and an Fe$_{0.5}$Mn$_{0.5}$TiO$_3$ Ising spin glass.
Differences in the nonequilibrium dynamics of the strongly interacting nanoparticle system and the two spin glass samples are discussed in terms of anisotropy and different timescales, due to the much longer microscopic flip time of a magnetic moment than of an atomic spin.


\newpage

\section{Introduction}
\label{Chap: intro}

Ferro- and ferrimagnetic nanoparticles are important examples of how a reduction in size changes the  properties of a magnetic material.
For small particles it is energetically favorable to avoid domain walls and form only one magnetic domain.
The magnetism of such single-domain particles has been an active field of research since the pioneering work of Stoner and Wohlfarth \cite{stowoh48} and N{\'e}el \cite{nee49} in the late 1940s.
Because of new fabrication methods and characterization techniques, understanding of and interest in nanosized materials have increased explosively within the disciplines of physics, chemistry, material science, and medicine.
This development is driven by a large number of applications; nanosized magnetic materials are used in, for example, magnetic recording media, ferrofluids, catalysts, and refrigerators; as well as by a large interest of fundamental nature.
Nanomagnets made up of a small number of spins can be used to study quantum tunneling of magnetization \cite{wernsdorfer2001}.
In ferrofluids the dipolar interparticle interaction can be tuned by the particle concentration, and 
frozen ferrofluids have been shown to change their magnetic behavior from superparamagnetic at low concentrations to spin-glass-like in dense systems.

The research on spin glasses started in the 1970s after the discovery by Cannella and Mydosh \cite{canmyd72} of a peak in the ac susceptibility of diluted gold--iron alloys.
Several different materials with various interaction mechanisms were soon found to exhibit this ``new'' magnetic behavior, all with two properties in common --- disorder and frustration.
Spin glasses have since been widely studied, partly because they are excellent model systems of materials with quenched disorder.
An understanding of spin glasses can thus contribute to the understanding of other, more complex disordered systems, such as ceramic superconductors, polymers, gels, and dense nanoparticle systems.

This article reviews the dynamic properties of magnetic nanoparticle systems with different interparticle interaction strength.
In section \ref{Chap: nano} we discuss basic properties of noninteracting particle systems, and thermodynamic perturbation theory is used to study weakly interacting particle systems.
In section \ref{Chap: nanovxv} we discuss the behavior of strongly interacting magnetic nanoparticle systems in the light of recent results in the field of spin glasses.

\section{Single-Domain Magnetic Nanoparticles}
\label{Chap: nano}


The study of single-domain magnetic particles has been an
active field of research since the pioneering work of Stoner and
Wohlfarth \cite{stowoh48}, who studied the hysteretic rotation of the
magnetization over the magnetic--anisotropy energy barrier under
the influence of an applied field, and N{\'e}el \cite{nee49}, who
predicted that at nonzero temperature the magnetization can
overcome the energy barrier as a result of thermal agitation.
Later, Brown \cite{bro63} derived the
Fokker--Planck equation for the probability distribution of spin
orientations, starting from the stochastic Landau--Lifshitz equation,
and calculated approximate expression for the relaxation time of particles with uniaxial anisotropy.
The theoretically most well studied systems are noninteracting classical spins (representing the magnetization of the nanoparticles) with axially symmetric magnetic anisotropy. 
A great step forward in comparing experiments and theory was taken when measurements on individual particles were reported \cite{weretal97TA}.
A profound knowledge of the physical properties of isolated particles is a prerequisite for further studies of phenomena such as quantum tunneling in molecular  nanomagnets or dipole--dipole interaction in dense samples.

In this section, some general properties of magnetic nanoparticles are first recalled (for more details, see, e.g., Refs. \cite{dorfiotro97,garpal2000acp,batlab2002}). 
The definition of a magnetic nanoparticle is rather wide and includes ferro- and ferrimagnetic materials (e.g., $\gamma$-Fe$_2$O$_3$, Fe$_3$O$_4$, and Fe$_{1-x}$C$_x$) as well as magnetic nanoclusters (e.g., Mn$_{12}$ and Fe$_8$). 
Subsequently, thermodynamic properties of spins weakly coupled by the dipolar interaction are calculated. 
Dipolar interaction is, due to its long range and reduced symmetry, difficult to treat analytically; most previous work on dipolar interaction is therefore numerical \cite{zalcie93,zal96,andetal97,bergor2001}. 
Here thermodynamic perturbation theory will be used to treat weak dipolar interaction analytically.
Finally, the dynamical properties of magnetic nanoparticles are reviewed with focus on how relaxation time and superparamegnetic blocking are affected by weak dipolar interaction.
For notational simplicity, it will  be assumed throughout this section  that the parameters characterizing different nanoparticles are identical (e.g., volume and anisotropy).

\subsection{General Properties}
The current  studies of magnetic single-domain nanoparticles are limited to systems where the particles are fixed in space (realized, e.g., in frozen ferrofluids, single crystals of molecular magnets, and magnetic nanoparticles in a solid matrix).
We will also assume that every single-domain nanoparticle is in internal thermodynamic equilibrium and that its constituent spins rotate coherently. 
Moreover, we are considering only temperatures much lower than the Curie temperature, so the spontaneous magnetization is approximately constant with temperature.
Hence, the only relevant degree of freedom is the {\em orientation} of the net magnetic moment.

The Hamiltonian of a single isolated nanoparticle consists of the magnetic anisotropy (which creates preferential directions of the magnetic moment orientation) and the Zeeman energy (which is the interaction energy between the magnetic moment and an external field).
In the ensembles, the nanoparticles are supposed to be well separated by a nonconductive medium (i.e., a ferrofluid in which the particles are coated with a surfactant).
The only relevant interparticle interaction mechanism is therefore the dipole--dipole interaction.

\subsubsection{Magnetic Anisotropy}
The term {\em magnetic anisotropy} is used to describe the dependence of the internal energy on the direction of the spontaneous magnetization of the ferro/ferrimagnetic nanoparticle, creating  ``easy'' and ``hard'' directions of magnetization.
In general, a bulk sample of a ferromagnet exhibits magnetic anisotropy with the same symmetry as in the crystal structure.
This anisotropy energy originates from spin--orbit coupling and is called {\it magnetocrystalline anisotropy} \cite{chikazumi}.
The two most common symmetries are uniaxial and cubic.
For uniaxial symmetry the energy is given by
\be
E_a^{\rm uni} = K_1 V \sin^2\theta +  K_2 V \sin^4\theta + \cdots  \, ,
\label{Eq: mae-uni}
\ee
where $V$ is the particle volume, $K_1$ and $K_2$ are anisotropy constants, and $\theta$ is the angle between the magnetic moment and the symmetry axis.
For cubic symmetry the anisotropy can be expressed in terms of the direction cosines ($\alpha_i$) as
\be 
E_a^{\rm cubic} 
= 
K_1 V(\alpha_1^2\alpha_2^2+\alpha_2^2\alpha_3^2+\alpha_3^2\alpha_1^2)
+
K_2 V\alpha_1^2\alpha_2^2\alpha_3^2 + \cdots  \, ,
\ee
where the $\alpha_i$ are defined through $\alpha_1=\sin\theta\cos\phi$, $\alpha_2=\sin\theta\sin\phi$ and $\alpha_3=\cos\theta$, $\theta$ is the angle between the magnetization and the $z$ axis, and $\phi$ is the azimuthal angle.

For a single-domain ferromagnet, any nonspherical particle shape gives rise to {\it shape anisotropy} due to the internal magnetostatic energy.
The magnetostatic energy, for an ellipsoid of revolution, is equal to
\be
E_m = \case{1}{2}\mu_0 V \Ms^2 (N_z \cos^2\theta + N_x \sin^2\theta),
\ee
where $\theta$ is the angle between the magnetic moment and the polar axis $\z$, $\Ms$ is the saturation magnetization, $N_z$ is the demagnetization factor along the polar axis, and $N_x = N_y$ is the demagnetization factor along an equatorial axis.
Both the magnetostatic energy for an ellipsoid and the uniaxial magnetocrystalline anisotropy energy [Eq. (\ref{Eq: mae-uni})] can, to first order --- except for a constant term --- be written as
\be
E_a = - \D \cos^2\theta \, ,
\label{Eq: Ea-uni}
\ee
where $\D=KV$ is the anisotropy energy barrier and the uniaxial anisotropy constant $K= \case{1}{2}\mu_0 \Ms^2 (N_x -N_z)$ in the case of shape anisotropy.
For a prolate ellipsoid, $K>0$ and the anisotropy is of ``easy axis'' type, since there exist two minima of the anisotropy energy along $\pm \z$ (the anisotropy axis).
For an oblate ellipsoid, $K<0$ and the anisotropy energy has its minimum in the whole $xy$ plane. In this case the anisotropy is of ``easy plane'' type. 

With decreasing particle size, the magnetic contributions from the surface will eventually become more important than those from the bulk of the particle, and hence {\it surface anisotropy} energy will dominate over the magnetocrystalline anisotropy and magnetostatic energies.
A uniaxial anisotropy energy proportional to the particle surface $S$
\be
E_a^{\rm surface} = K_s S \cos^2\theta \,
\ee
has been observed experimentally by ferromagnetic resonance \cite{gazetal98}. 

Hereafter, we will assume uniaxial anisotropy, of easy-axis type, given by Eq. (\ref{Eq: Ea-uni}) (if not otherwise indicated), since it is the simplest symmetry that contains the basic 
elements (potential minima, barriers) responsible for the important role of 
magnetic anisotropy in superparamagnets.
Experimental evidence for uniaxial anisotropy is given in Refs. \cite{gazetal98,upasrimeh2000}.

\subsubsection{Superparamagnetic Relaxation}
The uniaxial anisotropy energy creates two potential wells separated by the energy barrier $\D$.
The magnetic moment is subjected to thermal fluctuations and may undergo a Brownian-type rotation surmounting the potential barriers. 
This relaxation process was proposed and studied by N{\'e}el in 1949 \cite{nee49} and further developed by Brown in 1963 \cite{bro63}.
In the high potential barrier range, $\beta \D \gg 1$, where $\beta=1/\kT$, the characteristic time for the overbarrier rotation $\tau$ can approximately be written in the Arrhenius form
\be
\label{Eq: arrhenius}
\tau 
\simeq
\tau_0 \exp(\beta \D) \,,
\ee
where $\tau_0 \sim 10^{-9} - 10^{-12}$ s.
For observation times $\tobs$ much longer than the relaxation time, $\m$ maintains the thermalequilibrium distribution of orientations as in a classical paramagnet; however, because of the much larger magnetic moment than a single spin, this phenomenon was called {\it superparamagnetism} \cite{bealiv59}.
The condition of superparamagnetism ($\tobs \gg \tau$) corresponds to a temperature range that fulfills $\ln(\tobs / \tau_0) > \beta \D$ .
For $\tobs \sim 10$ s, due to the small value of $\tau_0$, this equilibrium range extends down to low thermal energies  compared to the anisotropy energy  ($25 > \beta \D$). 
Hence, {\em within the equilibrium regime}, the system displays an  isotropic behavior at high temperatures ($\beta \D  \ll 1$), but a  strongly anisotropic behavior at low temperatures ($\beta \D \gg 1$).

If $\tobs \ll  \tau$, the magnetic moment is {\it blocked} in one of the potential wells, a state that corresponds to stable magnetization in a bulk magnet.
If the measurement time is of the same order as the relaxation time ($\tobs \sim \tau$), dynamical time-dependent effects are observed.

\subsubsection{Effects of a Magnetic Field}

The Hamiltonian of a noninteracting nanoparticle with uniaxial anisotropy is given by
\be
\label{Eq: H-nonint}
\Hamil = -\frac{\D}{m^2}(\m \cdot \n)^2 - \mu_0 \m \cdot \HH
\ee
where $\m$ is the magnetic moment with $m = \Ms V$, $\D=KV$
and $\n$ is a unit vector along the symmetry axis of the anisotropy energy (anisotropy direction).
By introducing unit vectors for the magnetic moment ($\e = \m/m$) and the external magnetic field ($\h = \HH / H$) and defining dimensionless parameters for the anisotropy and magnetic field 
\be
\label{Eq: sigma-xi}
\sigma = \beta \D, \qquad \xi = \beta \mu_0 m H,
\ee
we can write a dimensionless Hamiltonian as
\be
- \beta \Hamil = \sigma (\e \cdot \n)^2 + \xi  (\e \cdot \h).
\ee

\begin{figure}
\begin{center}\leavevmode
\includegraphics[width=0.75\textwidth]{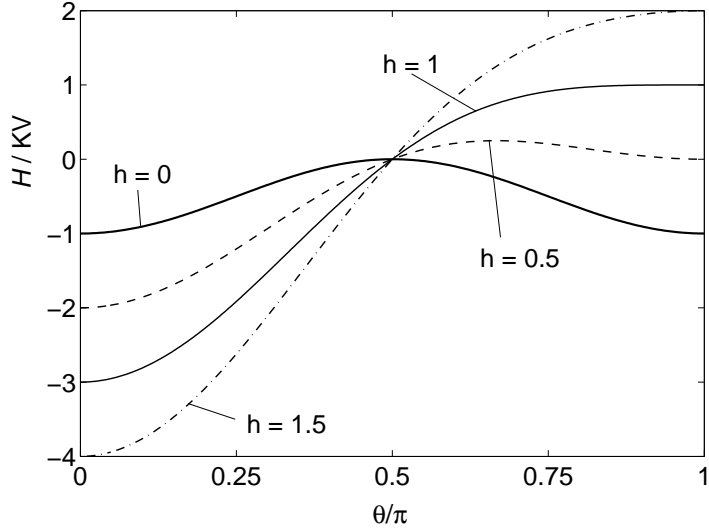}
\caption{Magnetic energy versus $\theta$ in the case of a longitudinal field for different values of the reduced field $h=H/H_K$.
\label{Fig: barrier}}
\end{center}
\end{figure}

The bistable character of the zero-field Hamiltonian will be destroyed by a sufficiently large field. The critical field for $\n \| \h$ is called the {\em anisotropy field}, and is expressed by
\be
 H_K = \frac{2 A}{\mu_0 m} = \frac{ 2 K}{\mu_0 \Ms}.
\ee  
We can define another dimensionless field quantity
\be
 h= \frac{H}{H_K}=\frac{\xi}{2\sigma} \; ,
\ee
which is the field measured in units of the anisotropy field.
The Hamiltonian, as a function of the angle $\theta$ between the anisotropy axis and the magnetic moment ($\e \cdot \n = \cos \theta$), is shown in Fig.~\ref{Fig: barrier} for different values of the longitudinal field.

\subsubsection{Interparticle Interaction}
\label{Sec: dipolar}
Dipole--dipole interaction is present in all  magnetic spin systems, but usually other interaction mechanisms such as exchange interaction dominate. 
The relative weakness of the dipolar coupling between magnetic ions in paramagnetic systems results in characteristic temperatures lying in the range of 0.01--0.1 K.
For superparamagnetic nanoparticles (for which care has been taken to avoid direct contact between the particles by, e.g., applying a surfactant to a ferrofluid), exchange interaction and other interaction mechanisms can usually be discarded so that the dipolar interaction is the only relevant interparticle interaction. 
In addition, the size of the typical magnetic moment ($S \sim 10^2 - 10^5$ magnetic spins) shifts the relevant temperatures up to the range of a few kelvins, making it possible to observe effects of dipolar interaction in conventional magnetization experiments.

The dipolar field, created by all other spins, at the position $\vec{r}_i$ of the spin $\e_i$, is given by
\be
\label{Eq: Hlocal}
\HH_i=\frac{m}{4\pi a^3} \sum_j \Gij \cdot \ej \, ,
\ee
where the term $j=i$ is omitted from the summation, $a$ is defined in such a way that $a^3$ is the mean volume around each spin, and
\begin{eqnarray}
\label{Gij}
\Gij
&=&
\frac{1}{r_{ij}^{3}}
\left(3\,\vij \vij - \I\right),
\\
\rij
&=&
\vec{r}_{i}-\vec{r}_j,
\quad
\vij
=
\frac{\rij}{r_{ij}}
\;,
\end{eqnarray}
where $\I$ is the unit tensor.

By introducing the dimensionless coupling constant
\begin{equation}
\xid
=
\frac{\mu_{0}m^2}{4\pi a^3}
\frac{1}{\kT}
\;,
\end{equation}
and noting that the dipolar energy $\Ed = \frac{\mu_0}{2} \sum_{i\ne j} \m \cdot \HH_i$, we can write the total dimensionless Hamiltonian of an interacting nanoparticle system as
\begin{equation}
-\beta \Hamil
=
\sigma \sum_{i}(\ei \cdot \nii)^2
+
\xi \sum_i (\ei \cdot \h)
+
\xid
\sum_{i>j}\wij
\;,
\end{equation}
where $\wij=\ei \cdot \Gij \cdot \ej$.
Note that the interaction strength can also be measured by the
temperature independent coupling parameter 
\begin{equation}
\label{Eq: hd}
\hd
=
\frac{\xid}{2\sigma}
=
\frac{\Ms }{4 \pi H_K}c
\;,
\end{equation}
which is the magnitude of the field, measured in units of the anisotropy field $H_{K}$, produced at a given position by a dipole at a distance $a$.
Here, $c=V/a^3$ is the volume concentration of particles.

Dipole--dipole interaction is long-ranged and anisotropic, which makes it cumbersome to treat both analytically and numerically.
Randomness in particle positions and anisotropy directions yields frustration and magnetic disorder leading to glassy dynamics for strongly interacting systems \cite{luoetal91}.
We will therefore use analytical and numerical treatment only for weakly interacting particle systems, while strongly interacting systems will be discussed in terms of collective spin glass behavior in section \ref{Chap: nanovxv}.

\subsection{Thermal Equilibrium Properties}
\label{Sec: teq}

The thermal equilibrium average of any quantity
$\obs(\e_{1},\ldots,\e_{N})$ is given by
\begin{equation}
\label{theeqpro}
\la\obs\ra
=
\frac{1}{\Z}
\int\!
\dW
\obs
\,
\exp (-\beta \Hamil)
\;,
\end{equation}
where $\Z = \int\!\dW\exp (-\beta \Hamil)$
is the partition function and $\dW=\prod_{i}d\Omega_{i}$, with the solid angle
$d\Omega_{i}
=
d^{2}\ei/2\pi$.
In the case of noninteracting spins, Eq.~(\ref{theeqpro}) has already  been solved analytically for different quantities and anisotropies (for a review see, e.g., Ref. \cite{garpal2000acp}). 
For isotropic spins, the magnetization is given by the Langevin function and the linear susceptibility follows a $1/T$ dependence, $\chi_{\rm iso}=\frac{1}{3} \beta \mu_0 m^2$, while the nonlinear susceptibility follows a $1/T^3$ dependence, $\chi_{\rm iso}^{(3)}=-\frac{1}{45} \beta^3 \mu_0^3 m^4$.
Magnetic anisotropy generally causes deviations from these well-known laws \cite{garpal2000acp,garpallaz97,raiste97,garjonsve2000} and 
also dipole--dipole interparticle interaction will cause deviations from the Langevin behavior \cite{hukluc2000,jongar2001PRB}.
It is important to know the nature of such deviations in order to avoid confusion with, for example  deviations due to quantum effects \cite{hanjohmor98,tejetal98,mamnakfur2002}.

In the case of dipole--dipole interaction, the calculation of any thermodynamic property becomes a many-body problem and approximations and/or numerical simulations are needed in order to solve Eq.~(\ref{theeqpro}). 
Here we will use thermodynamic perturbation theory to expand the Boltzmann distribution in the dipolar coupling parameter --- an approximation that is valid for weakly coupled spins.

The analysis presented in this section is restricted to spins with Hamiltonians having inversion symmetry [$\Hamil(\m)=\Hamil(-\m)$].
This assumption is valid for any kind of anisotropy or dipolar interaction, but if a bias field is applied in addition to the probing field, the condition of inversion symmetry breaks down.

\subsubsection{Thermodynamic Perturbation Theory for Weakly Interacting Superparamagnets
\label{Sec: TPT}
}
We will consider dipolar interaction in zero field so that the total Hamiltonian is given by the sum of the anisotropy and dipolar energies $\Hamil = \Ea + \Ed$. By restricting the calculation of thermal equilibrium properties to the case $\xid \ll 1$, we can use thermodynamical perturbation theory \cite{pie33,lanlif5} to expand the Boltzmann 
distribution in powers of $\xid$. 
This leads to an expression of the form \cite{jongar2001PRB}
\begin{equation}
\label{W:approx}
W
=
\Wa
\left(
1
+
\xid
F_{1}
+\case{1}{2}
\xid^{2}
F_{2}
+
\cdots
\right)
\;,
\end{equation}
where $F_{1}$ is linear in $\Ed$ and
$F_{2}$ is (up to) quadratic in $\Ed$, 
 while
\begin{equation}
\label{W:non-int}
\Wa
=
\Z_a^{-1}
\exp(-\beta\Ea)
\;,
\end{equation}
is the Boltzmann distribution of the noninteracting ensemble.
Expressions for $F_1$ and $F_2$ in are given in Appendix \ref{App: expTPT}.
By keeping all averages weighted with $\Wa$, the thermal-equilibrium quantities calculated with this method will be exact in the anisotropy and only perturbational in the dipolar interaction.
An ordinary high-temperature expansion corresponds to expanding
Eq.\ (\ref{W:non-int}) further in powers of $\beta$.

All results obtained below with the thermodynamic perturbation theory are limited to the case of axially symmetric anisotropy potentials (see Appendix \ref{App:alg}), and all explicit calculations are done assuming uniaxial anisotropy (see Appendix \ref{App: Sl}).

\subsubsection{Linear Susceptibility}
The equilibrium linear susceptibility is, in the absence of an external bias field, given by
\begin{equation}
\label{lin-susc}
\chi
=
\frac{\mu_{0}m^{2}}{\kT}
\frac{1}{N}
\big\langle s_{z}^2 \big\rangle
,
\qquad
s_{z}=\sum_{i}( \ei\cdot\h )
,
\end{equation}
where $\h$ is a unit vector along the probing field direction 
 and $s_{z}$ is the field projection of the net magnetic moment.
Calculating $\la  s_{z}^2 \ra$ using thermodynamic perturbation theory yields an expansion of the susceptibility of the form
\be
\label{xeq-TPT}
\chi
=
\frac{\mu_{0}m^{2}}{\kT} \left( a_0 + \xid a_1 + \case{1}{2} \xid^2 a_2  \right) \, ,
\ee
with the general expressions for the coefficients $a_n$ given in Appendix \ref{App:linsusc}.
Simplified expressions for the coefficients can be obtained for 
some orientational distributions of the anisotropy axes, such as parallel anisotropy axes and randomly distributed axes.

For systems with parallel axes (e.g., single crystals of magnetic
molecular clusters or a ferrofluid frozen in a strong field), the
coefficients for the longitudinal response read
\begin{eqnarray}
\label{a0para}
\coeff_{0,\parallel}
&=&
\frac{1+2S_{2}}{3} \,
\\
\label{a1para}
\coeff_{1,\parallel}
&=&
\frac{1+4S_{2}+4S_{2}^2}{9}
\,
\Qls \, ,
\\
\label{a2para}
\case{1}{2}
\coeff_{2,\parallel}
&=&
-\frac{1+4S_{2}+4S_{2}^2}{27}
\left[
(1-S_{2}) \, \big( \Rls - \Sls\big)
+
3S_{2} \, \left( \Tls - \Uls \right)
\right]
\\
& &
{}+ \frac{7 + 10S_{2}-35S_{2}^2+18S_{4}}{315}
\left[
(1-S_{2}) \, \Vls
+
3S_{2} \, \big(\Tls - \case{1}{3}\Rls\big)
\right]
\nonumber
\;,
\end{eqnarray}
where $\Qls$, $\Rns$ ($\Rls$), $\Sls$, $\Tls$ and  $\Uls$ are
{\em lattice sums} whose properties are discussed in Sec. \ref{Sec: latt}.
The properties of $S_l(\sigma)$ are discussed in Appendix \ref{App: Sl}.

A common experimental situation is an ensemble of nanoparticles with the anisotropy axes oriented randomly (e.g., frozen ferrofluids).
To obtain the susceptibility when the anisotropy axes are distributed
at random, we average the general expressions for the $a_{n}$
over $\n$, getting
\begin{eqnarray}
\label{a0ran}
\coeff_{0,{\rm rand}}
&=&
\frac{1}{3}
\\
\label{a1ran}
\coeff_{1,{\rm rand}}
&=&
\frac{1}{9}
\,
\Qls
\\
\label{a2ran}
\case{1}{2}
\coeff_{2,{\rm rand}}
&=&
-\frac{1}{27}
\,
\big( \Rls - \Sls \big)
+
\frac{1}{45}\,
(1-S_{2}^2)\, \Vls
\;.
\end{eqnarray}
Note that in the limit of isotropic spins (where $S_{l}\to0$), the
results for coherent axes and for random anisotropy duly coincide
and agree with ordinary high-temperature expansions.

It can easily be shown that $\coeff_{0,{\rm rand}}$ is independent of anisotropy for any type of anisotropy, not only axially symmetric \cite{garjonsve2000}.
Changing the coordinate system of  \eq{lin-susc} to the local one determined by the anisotropy direction of each spin, the field becomes a randomly distributed vector and by performing random averaging with respect to $\h$ by means of \eq{alg2:iso}, one obtains
\be
\chi
=
\frac{\mu_{0}m^{2}}{\kT} \frac{1}{3N} \la \sum_{i}( \ei\cdot\ei ) \ra 
= \frac{\mu_{0}m^{2}}{3\kT} \;,
\ee
which is same expression as for isotropic spins.
It is, however, only for the linear susceptibility term that randomly distributed anisotropy axes erase all traces of the anistropy. 
For the nonlinear susceptibility, the anisotropy is of importance even for systems with randomly distributed anistropy axis \cite{garjonsve2000,garpallaz97,garpal2000acp,raiste97}.

\subsubsection{Specific Heat}

The specific heat at constant volume can be obtained directly from the partition function
\begin{equation}
\frac{c_{v}}{\kB}
=
\beta^2\frac{\partial^2}{\partial \beta^2}(\ln \Z)
=
\sigma^2\frac{\partial^2}{\partial \sigma^2}(\ln \Z)
\;,
\label{cV}
\end{equation}
where, to take the $\sigma$ derivative, the coupling parameter
$\xid$ is expressed as $\xid=2\sigma\hd$ [Eq.\ (\ref{Eq: hd})].
As in the calculation of $\chi$, we consider only the zero-field specific heat. 
In that case, the term linear in $\xid$ vanishes and the expansion of the specific heat to second order in $\xid$ reads
\begin{equation}
\label{cVexp}
\frac{c_{v}}{N \kB}=\sigma^2 b_0 + \case{1}{2} \xid^2 b_2
\;,
\end{equation}
where the zeroth-order coefficient
\begin{equation}
\label{b0gen}
b_0
=
\frac{4}{315}(18S_4 -35S_2^2+10S_2+7)
\;,
\end{equation}
gives the specific heat in the absence of interaction \cite{garpal2000acp}.
The general formula for $b_{2}$ is given in Appendix \ref{App:cV}.
Again, it is possible to obtain simplified formulae for coherent anisotropy axes and for random anisotropy.
In the first case ($\nii=\n$, $\forall i$), we obtain
\begin{eqnarray}
b_{2,\parallel}
&=&
\case{1}{3}
\Big\{
1-S_2^2
-
4\sigma S_2 S_2' 
-
\sigma^2 [S_2 S_2'' + (S_2')^2]
\Big\} 
\Rns
\nonumber \\
& &
+\case{1}{3}
\Big(
2S_2(1-S_2)
+
4\sigma S_2'(1-2S_2) 
\nonumber\\
& &
\qquad
{}+\sigma^2 \{S_2''-2[S_2 S_2'' + (S_2')^2]\}
\Big)
\Vls
\nonumber \\
&&+
\Big\{
S_2^2
+
4 \sigma S_2 S_2'
+
\sigma^2 [S_2S_2'' + (S_2')^2]
\Big\}
\Tls
\;,
\end{eqnarray} 
where $S_2'=dS_2/d\sigma$.
For randomly distributed axes, on averaging the general expression for
$b_2$ over $\n$, one simply gets
\begin{equation}
b_{2,{\rm rand}}=\case{1}{3}\cal{R}.
\end{equation}
This is the same correction term as that obtained for {\em isotropic} spins by Waller \cite{wal36} and van Vleck \cite{vanvle37} using ordinary high-temperature expansions.

\subsubsection{Dipolar Fields}
\label{sec:df}
We are interested in calculating thermodynamical averages of the dipolar field, to introduce them in the expression for the relaxation rate in a weak but arbitrary oriented field, in order to obtain an expression for the relaxation rate of weakly interacting dipoles (we will argue  in Sec. \ref{Sec: relax-int} that the effect of weak dipolar interaction can be accounted for by the thermodynamic averages of the dipolar field).
Because of the inversion symmetry of the anisotropy, only the square of the field will enter the low-field expression for the relaxation rate and not the field itself. 
In addition, the effects of longitudinal and transversal fields will be different.
To be able to calculate the dipolar field is also of interest in the study of quantum tunneling of the magnetization of molecular magnets (e.g., Fe$_8$ and Mn$_{12}$) \cite{wernsdorfer2001,prosta98}

The general expressions for  $\langle \xi_{i,\|}^2 \rangle $ and $\langle \xi_{i,\perp}^2 \rangle = \langle \xi_{i}^2 \rangle - \langle \xi_{i,\|}^2\rangle  $
to second order in $\xid$ are given in Appendix \ref{App:df}. 
If we consider infinite systems, the index $i$ can be removed since all spins have the same surroundings.
For a system with aligned anisotropy axes, the averaged fields
are given in terms of the lattice sums by
\bea
\label{h:para:perp}
\big\langle
\xi_{\|}^{2}
\big\rangle
&=&
\frac{\xid^{2}}{3}
\left[
\left(1-S_{2}\right)
\Rls
+
3S_{2}\,{\cal T}
\right],
\nonumber
\\
\big\langle
\xi_{\perp}^{2}
\big\rangle
&=&
\frac{\xid^{2}}{3}
\left[
3 \Rns - (1- S_2) \Rls
+ 3S_{2}\,(\Rns  -\Rls - {\cal T})
\right]
\;,
\eea
while for randomly distributed anisotropy axes they read
\begin{equation}
\label{h:para:random}
\overline{
\big\langle
\xi_{\|}^{2}
\big\rangle
}
=
\frac{\xid^{2}}{3}
{\cal R} \, ,
\qquad
\overline{
\big\langle
\xi_{\perp}^{2}
\big\rangle
}
=
\frac{\xid^{2}}{3}
2{\cal R}
\,.
\end{equation}
Note that the result for random anisotropy is identical to the result for isotropic dipoles.

\subsubsection{The Lattice Sums}
\label{Sec: latt}
An essential element of the expressions derived for $\chi$, $c_{v}$, and the dipolar fields are the following ``lattice sums:''\footnote{$\h$ should be replaced by $\n$ in the formulae for $c_{v}$ and the dipolar fields.} 
\begin{eqnarray}
\label{Qsum}
\Qls
&=&
\frac{1}{N}\sum_{i} \sum_{j \ne i}
\h\cdot\Gij\cdot\h
\\
\label{Rsum}
\Rns
&=&
\frac{2}{N}\sum_{i} \sum_{j \ne i}
r_{ij}^{-6}
\\
\label{Rpsum}
\Rls
&=&
\frac{1}{N}\sum_{i} \sum_{j \ne i}
\h\cdot\Gij\cdot \Gij\cdot\h
\\
\label{Ssum}
\Sls
&=&
\frac{1}{N}\sum_{i} \sum_{j \ne i} \sum_{k \ne j}
\h\cdot\Gij\cdot\Gjk\cdot\h
\\
\label{Tsum}
\Tls
&=&
\frac{1}{N}\sum_{i} \sum_{j \ne i}
(\h\cdot\Gij\cdot\h)^2
\\
\label{Usum}
\Uls
&=&
\frac{1}{N}\sum_{i} \sum_{j \ne i} \sum_{k \ne j}
(\h\cdot\Gij\cdot\h)(\h\cdot\Gjk\cdot\h)
\end{eqnarray}

The long range of the dipole--dipole interaction leads to a sample shape\footnote{Sample shape refers to the shape of the whole ensemble of nanoparticles {\em not} to the shape of the individual particles. The linear susceptibility exhibits a sample shape dependence, while the zero-field specific heat and the dipolar fields do not.}  dependence of the physical quantities dependent on an external magnetic field \cite{gri68,banetal98}.
In the expressions obtained for the susceptibility, this sample shape dependence is borne by the slowly convergent lattice sums $\Qls,\; \Sls$, and $\Uls$.
If we consider ``sufficiently isotropic'' lattices, in the sense of fulfilling
$\sum(r_x)^n=\sum(r_y)^n=\sum(r_z)^n$, such as cubic and completely
disordered lattices (incidentally, the type of arrangements for which
in the classical Lorentz cavity--field calculation the contribution of
the dipoles inside the ``small sphere'' vanishes) these lattice sums vanish for large spherical samples.
The sums $\Rns,\; \Rls$, and $\Tls$,  on the other hand, contain $r_{ij}^{-6}$, which make them rapidly convergent and sample shape-independent.
For sufficiently symmetric lattices $\Rls = \Rns = 16.8, \; 14.5, \; 14.5$ for simple cubic (sc), body-centered cubic (bcc), and face-centered cubic (fcc) structures, respectively \cite{pople53}, and $\Tls=13.54$ (sc), 3.7 (bcc), 4.3 (fcc).
Note that since $\Rls-\Rns=0$, some terms vanish in the above expressions for $\chi$, $c_{v}$, and the dipolar fields.

\paragraph{Sample Shape and Anisotropy Dependence}

\begin{figure}
\begin{center}
\includegraphics[width=0.75\textwidth]{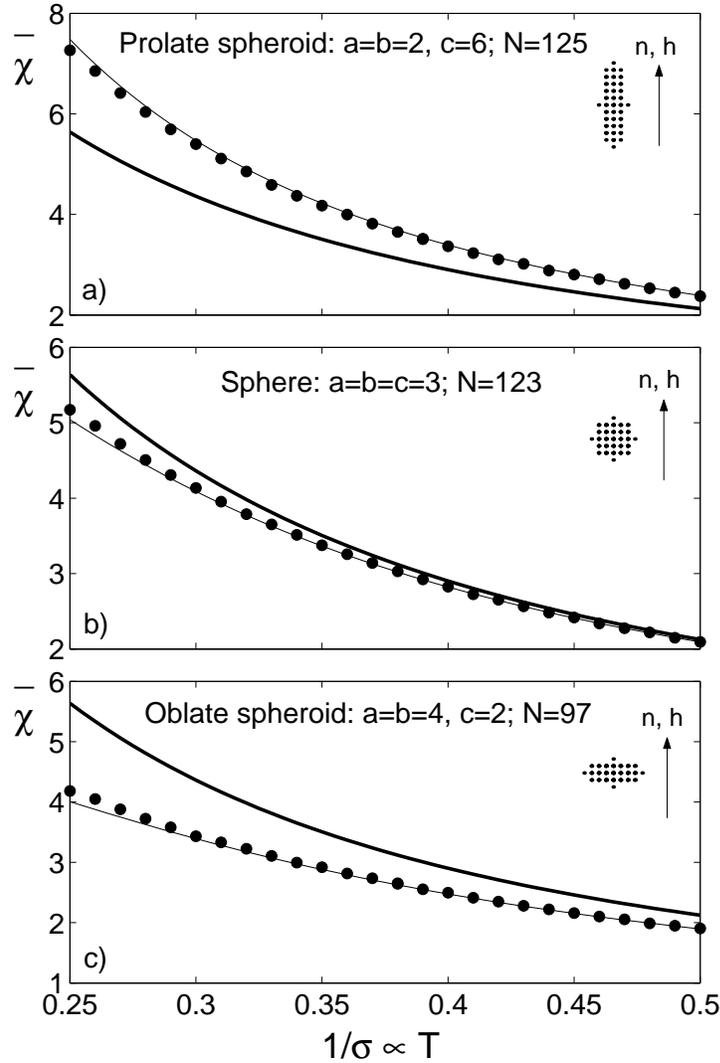}
\caption{Equilibrium linear susceptibility in reduced units $\bar{\chi} = \chi(H_K/m)$ versus temperature for three different ellipsoidal systems with equation $x^2/a^2 + y^2/b^2 + z^2/c^2 \leq 1$, resulting in a system of $N$ dipoles arranged on a simple cubic lattice. The points shown are the projection of the spins to the $xz$-plane. The probing field is applied along the anisotropy axes, which are parallel to the $z$ axis. 
The thick lines indicate the equilibrium susceptibility of the corresponding noninteracting system (which does not depend on the shape of the system and is the same in the three panels); thin lines show the susceptibility including the corrections due to the dipolar interaction obtained by thermodynamic perturbation theory [\eq{xeq-TPT}]; the symbols represent the susceptibility obtained with a Monte Carlo method. The dipolar interaction strength is $\hd=\xid/2\sigma = 0.02$.
\label{fig: Xeq-int}}
\end{center}
\end{figure}

The shape dependence of the linear susceptibility is illustrated in Fig. \ref{fig: Xeq-int}. The susceptibility is calculated for small systems using both thermodynamic perturbation theory and a Monte Carlo technique \cite{nowetal2000}, taking the dipolar interaction into account without any approximation.
It can be seen that $\chi$ obtained by thermodynamic perturbation theory accurately  describes the simulated susceptibility at high temperatures, while the results start to deviate at the lowest temperatures displayed.
An estimate of the lowest temperature attainable by the thermodynamic perturbation theory is $\xid \sim 1/6$, which is milder than the a priori restriction $\xid \ll 1$.

\begin{figure}[t]
\begin{center}
\includegraphics[width=0.75\textwidth]{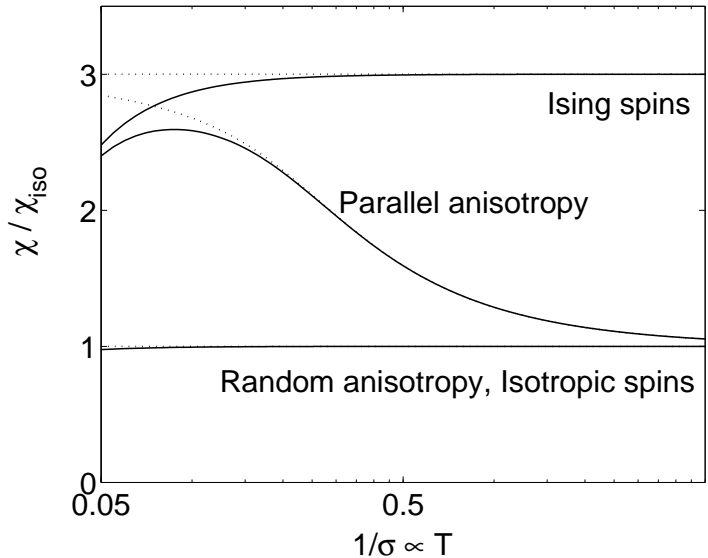}
\caption{Equilibrium linear susceptibility ($\chi/ \chi_{\rm iso}$) versus temperature for an infinite spherical sample on a simple cubic lattice.  The dotted lines are the results for independent spins, while the solid lines show the results for parallel and random anisotropy calculated with thermodynamic perturbation theory, as well as for Ising spins calculated with an ordinary high-temperature expansions. We notice in this case that the linear susceptibility for systems with random anisotropy is the same as for isotropic spins calculated with an ordinary high-temperature expansion.  
The dipolar interaction strength is $\hd=\xid/2\sigma = 0.004$. 
\label{fig: Xeq_ani}}
\end{center}
\end{figure}

For the linear susceptibility, the zero-field specific heat as well as the dipolar fields, the anisotropy dependence cancels out in the case of randomly distributed anisotropy axes (at least for sufficiently symmetric lattices). 
In other cases the anisotropy is a very important parameter as shown for the linear susceptibility in Fig.~\ref{fig: Xeq_ani} for an infinite (macroscopic) spherical sample.
The susceptibility is divided by $\chi_{\rm iso}=\frac{1}{3} \beta \mu_0 m^2$ in order to single out effects of anisotropy and dipolar interaction.
It can be seen in this figure that the effect of the dipolar interaction is much stronger for a system with parallel anisotropy axesthan for a system with random anistropy (which coincide with the case of isotropic spins). 
At low temperatures, the susceptibility of a system with parallel anisotropy  approaches that of Ising spins (calculated with an ordinary high-temperature expansion), while at high temperatures it is close to that of isotropic spins.

\begin{figure}[ht]
\begin{center}
\includegraphics[width=0.75\textwidth]{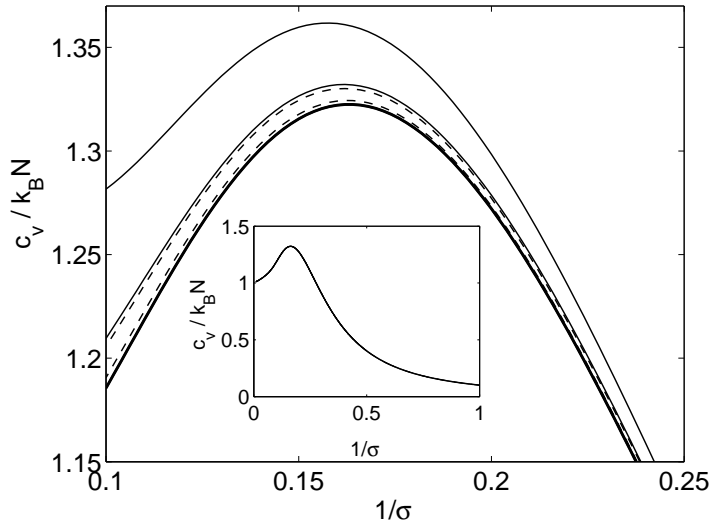}
\caption{
The specific heat per spin versus temperature for noninteracting spins
(thick line) and weakly interacting spins with randomly distributed
anisotropy axes (dashed lines) and parallel axes (thin lines) arranged
on a simple cubic lattice.
In each case, $\hd=\xid/2\sigma=0.003$ and $0.006$ from bottom to top.
The inset shows the specific heat for noninteracting spins over a
wider temperature interval.
\label{Fig:cv}}
\end{center}
\end{figure}

The specific heat for uncoupled spins does not depend on the orientations of the anisotropy axes; however, the corrections due to the dipolar coupling do, as can be seen in Fig. \ref{Fig:cv}). 
As for the linear susceptibility, the effect of dipolar interaction is stronger in the case of parallel anistropy than for random anistropy.

\subsection{Dynamic  Properties}

At high temperatures, a nanoparticle  is in a superparamagnetic state with thermal equilibrium properties as described in the previous section.
At low temperatures, the magnetic moment is blocked in one potential well with a small probability to overcome the energy barrier, while at intermediate temperatures, where the relaxation time of a spin is comparable to the observation time, dynamical properties can be observed, including magnetic relaxation and a frequency-dependent ac susceptibility.

For applications, such as magnetic recording media, it is necessary to know how different parameters will affect the relaxation time in order to avoid spontaneous data erasure (caused by  thermal fluctuations) on the lifetime of the device.
Because of the ongoing effort to increase the information/volume ratio, it is of special importance to know how the dipolar interaction of densely packed nanoparticles will affect the relaxation time.

\begin{figure}
\begin{center}
\includegraphics[width=0.75\textwidth]{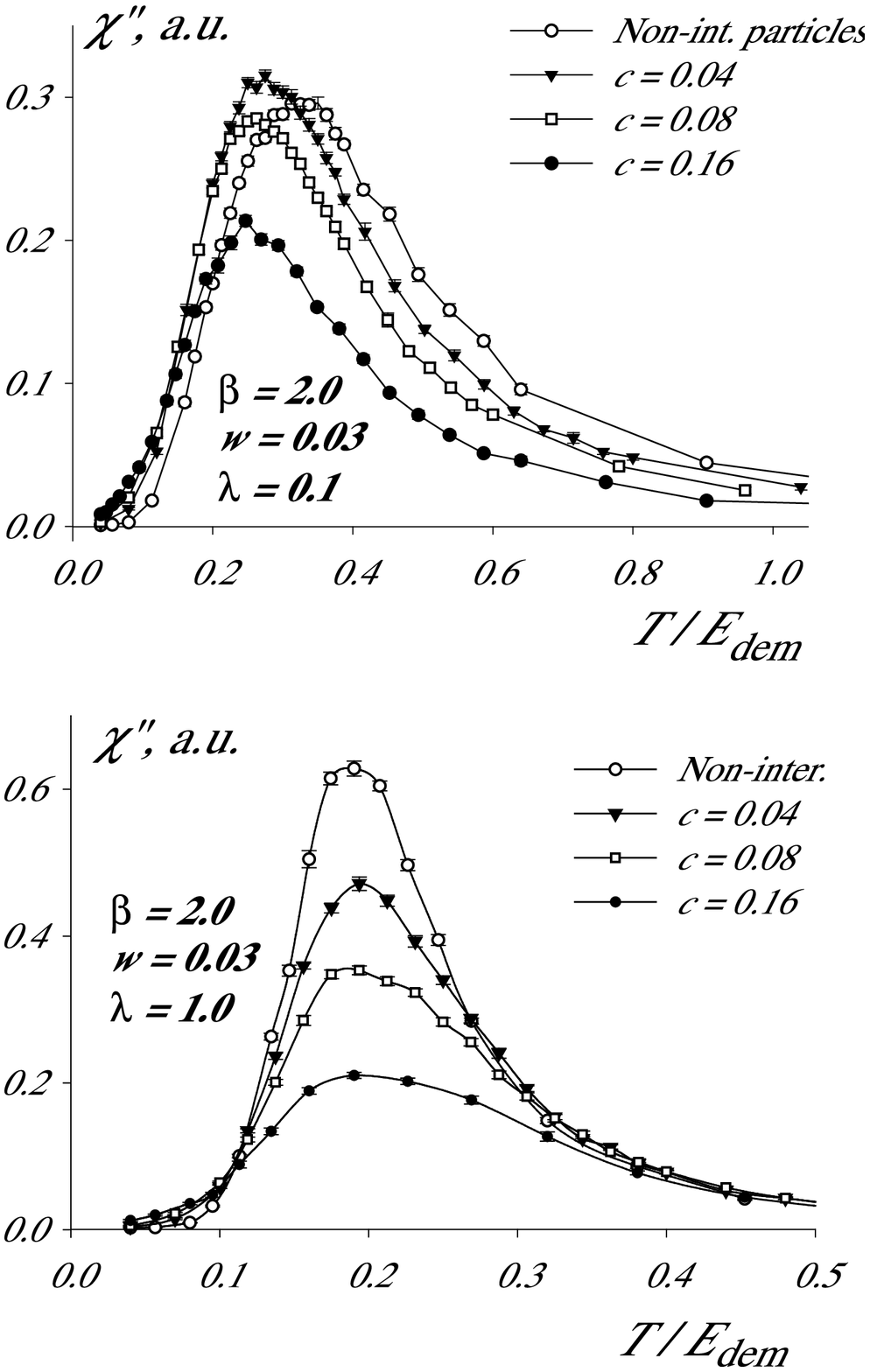}
\caption{The out-of-phase component of the ac susceptibility versus temperature for two different values of the damping $\lambda = 0.1$ and 1.0. $ \gamma \Ms \omega = 0.03$, $E_{\rm dem}= \Ms^2 V/\kB$, and $\hd = (\mu_0/4\pi)(c/\beta)$ with $\beta = 2.0$.  
From Ref. \cite{bergor2001}.
\label{Fig: berkov}}
\end{center}
\end{figure}

In 1963 Brown \cite{bro63} derived the Fokker--Planck equation for the probability distribution of the spin orientation, starting from the stochastic Gilbert equation, and calculated the relaxation time for particles with uniaxial anisotropy in a longitudinal field. More recent work on spins with nonaxially symmetric potential revealed a large dependence of the relaxation time on the damping coefficient $\lambda$ in the medium--weak damping regime \cite{cofetal98prb,cofetal98jpcm,garetal99}. 
Experiments on individual nanoparticles analyzed with accurate asymptotes of the relaxation time \cite{cofetal98prl} gave damping coefficients in the range $\lambda \approx 0.05 -0.5$.
Nonaxially symmetric potentials are, for example, created when applying a field at an oblique angle to a uniaxial spin. This oblique field can either be a bias field \cite{cofetal2001} or a nonlinear probing field \cite{garsve2000}.
In the case of interacting particles a transverse field component arises from the dipolar field of the surrounding particles. This explains the dependence on $\lambda$ of the blocking temperature that was first observed in numerical simulations by Berkov and Gorn \cite{bergor2001} (see Fig. \ref{Fig: berkov}).

The importance of including the damping in models describing the dynamic response of spins with nonaxially symmetric potentials (e.g., interacting uniaxial spins) tells us that models based only on how the energy barriers change \cite{dorbesfio88,mortro94} necessarily overlook 
the precession of the magnetic moment ($\lambda \to \infty$) and therefore 
cannot account for the numerical results of Berkov and Gorn. 
 A simple approach to include the damping in the dynamics of weakly coupled spin systems was proposed in Ref. \cite{jongar2001EPL}.

\subsubsection{The Equation of Motion}
We will present the equation of motion for a classical spin (the magnetic moment of a ferromagnetic single-domain particle) in the context of the theory of stochastic processes. 
The basic Langevin equation is the stochastic Landau--Lifshitz(--Gilbert) equation \cite{bro63,kubhas70}.
More details on this subject and various techniques to solve this equation can be found in the reviews by Coffey et al. \cite{cofcrekal93} and Garc{\'{\i}}a-Palacios \cite{garpal2000acp}.

\paragraph{Deterministic equations}
The motion of a magnetic moment can be described by the Gilbert equation \cite{gil55}
\be
\label{Eq: gilbert}
\frac{1}{\gamma}\frac{d\m}{dt}=\m \times \Beff - \frac{\lambda}{\gamma m} \m \times \frac{d\m}{dt}
\ee
where $\gamma$ is the gyromagnetic ratio (which includes $\mu_0$), $\lambda$ is a dimensionless damping coefficient, and the effective field is given by
\be
\Beff = - \frac{\mu_0^{-1}}{\partial\Hamil/\partial \m} \, .
\ee
The first term on the right side of \eq{Eq: gilbert} represents the precession of the magnetic moment about the axis of the effective field, while the second one is the damping term, which rotates $\m$ toward the potential minima and is responsible for the dissipation of the energy.

The Gilbert equation can be cast into the Landau--Lifshitz form \cite{lanlif35}
\be
\frac{1}{\gamma} \frac{d\m}{dt}=\m \times \Beff 
- \frac{\lambda}{m} \m \times (\m \times \Beff )
\ee
with a ``renormalized'' gyromagnetic ratio $\gamma \to \gamma/(1+\lambda^2)$ \cite{cofcrekal93,garpal2000acp}. 
In the case of uniaxial anisotropy and a Hamiltonian given by \eq{Eq: H-nonint},
$\Beff = (H_K/m)(\m \cdot \n)\n + \HH$, where $H_K$ is the anisotropy field and $\HH$ is an external field.
The ferromagntic resonance frequency $\omega$ for the precession about $\Beff$ is given by $\omega = \gamma \mu_0 H_{\rm eff}$ \cite{raiste94PRB}.

\paragraph{Stochastic Equations}
At $T \neq 0$ the magnetic moment will interact with the microscopic degrees of freedom (phonons, conducting electrons, nuclear spins, etc.). 
The complexity of this interaction allows an idealization, namely, to introduce them through a stochastic model. The simplest model is the Brownian, in which the interaction of $\m$ with the surroundings is represented by a randomly fluctuating magnetic field.
This fluctuating field is necessarily combined with a dissipation (damping) term, and these two terms are linked by fluctuation--dissipation relations \cite{kub66}.

In the work of Brown \cite{bro63} and Kubo and Hashitsume \cite{kubhas70} the starting equation is the Gilbert equation (\ref{Eq: gilbert}), in which the effective field is increased by  a fluctuating field yielding
the stochastic Gilbert equation. This equation can, as in the deterministic case, be cast into the Landau--Lifshitz form as
\be
\frac{1}{\gamma} \frac{d\m}{dt}=\m \times [\Beff + \bfl(t) ]
- \frac{\lambda}{m} \m \times \{\m \times [\Beff +\bfl(t)]\} \, ,
\label{Eq: LLG}
\ee
known as the stochastic Landau--Lifshitz--Gilbert (LLG) equation.
The fluctuating field is assumed to be Gaussian distributed white noise
\be
\langle b_{{\rm fl},\alpha}(t)\rangle = 0, \qquad 
\langle b_{{\rm fl},\alpha}(t) b_{{\rm fl},\beta}(t')\rangle =
2D \delta_{\alpha \beta} \delta(t-t') \, ,
\ee
with $\alpha,\beta={x,y,z}$. 
Garc{\'{\i}}a-Palacios and L{\'a}zaro \cite{garpallaz98} showed that the stochastic Landau--Lifshitz--Gilbert [Eq.~(\ref{Eq: LLG})] and the simpler stochastic Landau--Lifshitz (LL) equation,
\be
\frac{1}{\gamma} \frac{d\m}{dt}=\m \times [\Beff + \bfl(t) ]
- \frac{\lambda}{m} \m \times (\m \times \Beff) \, ,
\ee
both give rise to the same Fokker--Planck equation, describing the average properties of the magnetic moment, but with different Einstein-type relations between the amplitude of the fluctuating field and the temperature:
\be
D_{\rm LLG} = \frac{\lambda}{1+\lambda^2}\frac{\kT}{\gamma m}, \qquad
D_{\rm LL} = \lambda \frac{\kT}{\gamma m} \, .
\ee

\subsubsection{Relaxation Time in a Weak but Arbitrary Field}
\begin{figure}
\begin{center}
\includegraphics[width=0.75\textwidth]{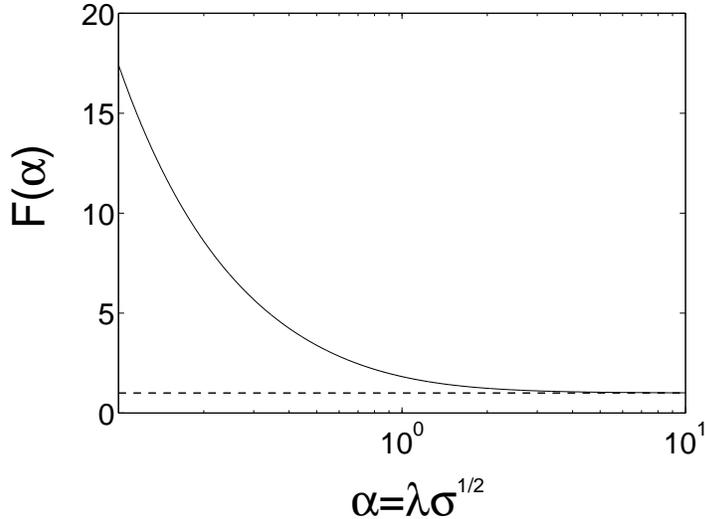}
\caption{$F$ versus $\alpha = \lambda \sigma^{1/2}$ (solid line) and the overdamped value $F=1$ (dashed line). 
\label{Fig: F}}
\end{center}
\end{figure}

We are interested in knowing how the relaxation time of uniaxial spins is affected by a weak field at an arbitrary direction, since it will allow us to study how the superparamagnetic blocking is affected by a field.
This field dependence of the relaxation time can be obtained by expanding the relaxation rate  $\Gamma = 1 / \tau$ in powers of the field components. As the spins have inversion symmetry in the absence of a field, $\Gamma$ should be an even function of the field components, and to third order it is given by 
\be
\label{Gamma:expansion:uni:gen}
\Gamma
\simeq
\Gamma_{0}
\Big(
1
+
c_{\|}
\xi_{\|}^{2}
+
c_{\perp}
\xi_{\perp}^{2}
\Big)
\;,
\end{equation}
where $\xi_{\|}$ and $\xi_{\perp}$ respectively are the longitudinal and transversal components of the field [given in temperature units, see \eq{Eq: sigma-xi}] with respect to the anisotropy axis.
$\Gamma_0$ is the zero-field relaxation rate, which for low temperatures ($\sigma >1$) is given by Brown's result \cite{bro63}
\be
\Gamma_{0}
=
\frac{1}{\tD}
\frac{2}{\sqrt{\pi}}
\sigma^{3/2}
e^{-\sigma}
\;,
\ee
where $\tD=m/(2\gamma \lambda \kT)$ is the relaxation time of isotropic spins.
The coefficient $c_{\|}$ can be obtained by expanding the expression for $\Gamma$ in the presence of a longitudinal field \cite{bro63,aha69}
\bea
\Gamma(\xi_{\|},\xi_{\perp}=0)
&
=
&
\frac{1}{\tD}
\frac{\sigma^{3/2}}{\sqrt{\pi}}
[(1+h)e^{-\sigma(1+h)^2}+(1-h) e^{-\sigma(1-h)^2}]
\nonumber
\\
&
\simeq
&
\Gamma_{0}
\Big(
1
+
\case{1}{2}
\xi_{\|}^{2}
\Big) \; .
\eea
There is no general expression for the relaxation rate in the presence of a nonzero transversal field valid for all values of the relevant parameters \cite{cofetal2000}, but Garanin et al. have derived a low-temperature formula valid for weak transversal fields \cite{garetal99}, which can be used to determine the coefficient $c_{\perp}$:
\bea
&&
\Gamma(\xi_{\|}=0,\xi_{\perp})
\simeq
\Gamma_{0}
\Big[
1
+
\case{1}{4}
F(\alpha)
\xi_{\perp}^{2}
\Big]
\,,
\\
&&
F(\alpha)
=
1
+
2
(2\alpha^{2}e)^{1/(2\alpha^{2})}
\gamma
\left(
1+\frac{1}{2\alpha^{2}}
\,,\,
\frac{1}{2\alpha^{2}}
\right)
\,.
\eea
Here $\alpha=\lambda\,\sigma^{1/2}$ and
$\gamma(a,z)
=
\int_{0}^{z} dt\, t^{a-1}\,e^{-t}$
is the incomplete gamma function.
It can be noted that for the axially symmetric potential with a longitudinal field, the only dependence on $\lambda$ is the trivial one in $\tD$, while in the nonaxially symmetric potential obtained with a transversal field the relaxation rate will strongly depend on $\lambda$ through $F(\alpha)$, which is plotted in Fig.~\ref{Fig: F}.

Collecting these results, we can finally write the expression for the relaxation rate in a weak field \cite{jongar2001EPL}: 
\begin{equation}
\label{Gamma:expansion:uni}
\Gamma
\simeq
\Gamma_{0}
\Big[
1
+
\case{1}{2}
\xi_{\|}^{2}
+
\case{1}{4}
F(\alpha)
\xi_{\perp}^{2}
\Big]
\;.
\end{equation}

\subsubsection{Relaxation Time of Weakly Interacting Nanoparticles}
\label{Sec: relax-int}
The relaxation time for weakly interacting nanoparticles with uniaxial anisotropy can be obtained by inserting the thermodynamical averages of the dipolar fields (calculated in section \ref{sec:df}) in the expression for the relaxation rate in a weak field [\eq{Gamma:expansion:uni}], yielding 
$ \Gamma \simeq \Gamma_{0}[ 1 + \case{1}{2} \la \xi_{\|}^{2}\ra + \case{1}{4} F(\alpha)
\la \xi_{\perp}^{2} \ra]$. 
Other models \cite{dorbesfio88,mortro94,luietal2002} are energy-barrier-based and therefore neglect the  $\lambda$ dependence on relaxation time. 
For instance, the model by M{\o}rup and Tronc \cite{mortro94} is basically the same as the one presented here in the particular case of  high damping and random anisotropy.
The  M{\o}rup--Tronc model predicts a decrease of the blocking temperature with increasing interaction strength for weak interaction as was observed in high-frequency M{\"o}ssbauer experiments, while the Dormann--Bessias--Fiorani model \cite{dorbesfio88} predicts an increase of the blocking temperature with increasing interaction strength as commonly observed in magnetization measurements.
These discrepancies led to some controversy \cite{hanmor98,dorfiotro99}.
Berkov and Gorn \cite{bergor2001} showed, by numerical integration of the stochastic LLG equation [\eq{Eq: LLG}], that for strong anisotropy (or weak interaction) the blocking temperature decreases with interaction (see Fig. \ref{Fig: berkov}), while for weak anisotropy (or moderate--high interaction) the energy barriers are governed by the interaction and hence grow with $h_d$.
An increase of the apparent blocking temperature is clearly the case for the strongly interacting nanoparticle samples (see Section \ref{Chap: nanovxv}) in which the relaxation time increases with $h_d$ due to spin--spin correlations (see Fig. \ref{Fig: tau}).

\begin{figure}[ht]
\begin{center}
\includegraphics[width=0.75\textwidth]{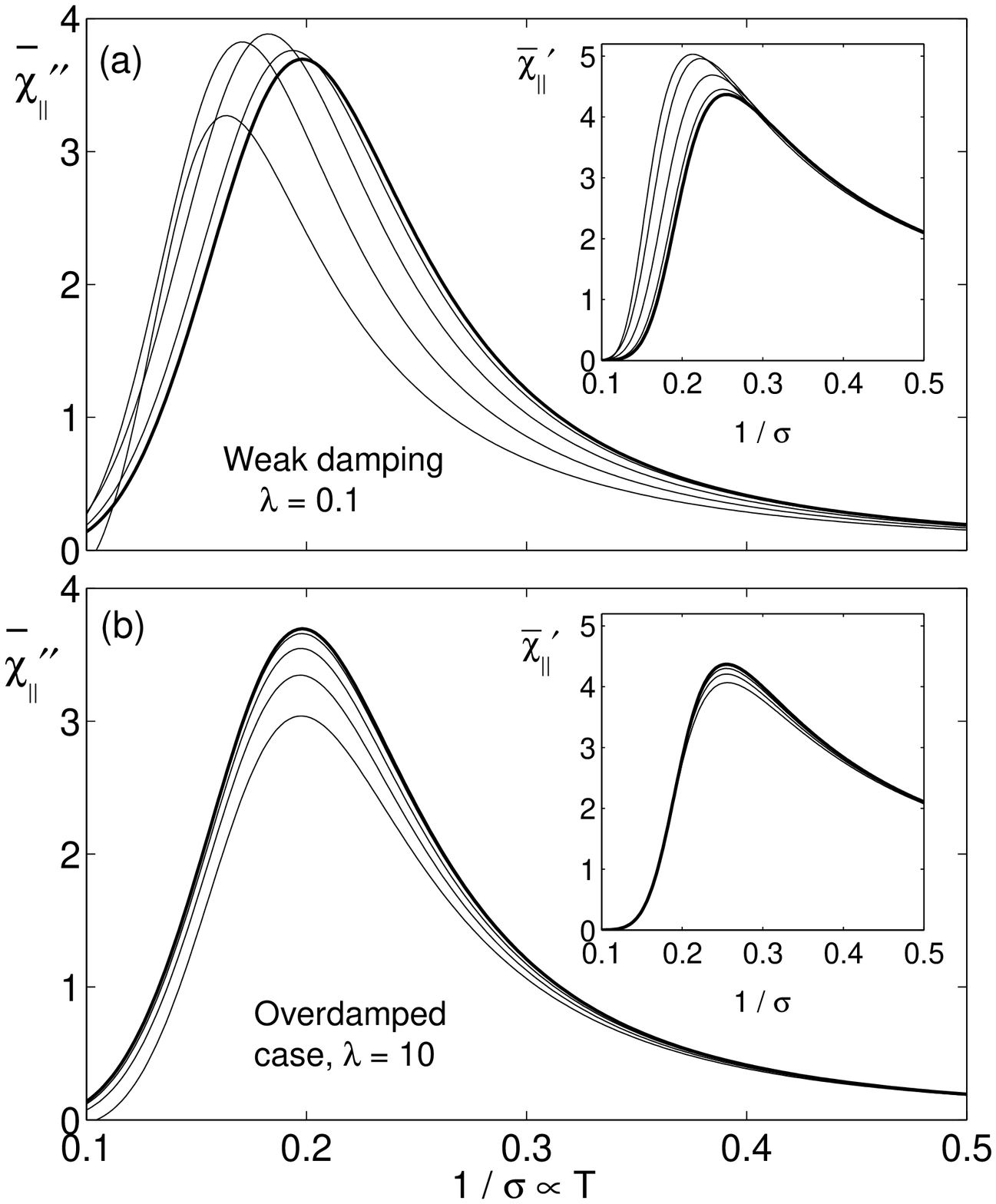}
\caption{
Imaginary component of the dynamical susceptibility versus
temperature (the real component is shown in the inset) 
for a spherical sample and spins placed in a simple cubic lattice.
The anisotropy axes are all parallel, and the response is probed along
their common direction.
The dipolar interaction strength $\hd=\xid/2\sigma$ is $\hd=0$
(thick lines), 0.004, 0.008, 0.012, and 0.016 from (a) right to left and
(b) top to bottom.
The frequency is $\omega\tD/\sigma=2\pi\times0.003$.
}
\label{Fig: ac-interacting}
\end{center}
\end{figure}

In order to determine the characteristics of the superparamagnetic blocking we use the equilibrium susceptibility $\chieq$ calculated using the thermodynamic perturbation theory \eq{xeq-TPT} and the relaxation rate $\Gamma$ obtained when the dipolar fields [\eq{h:para:perp} or (\ref{h:para:random})] are introduced in \eq{Gamma:expansion:uni}. 
Combining these expressions in a Debye-type formula
\be
\chi = \chieq \frac{\Gamma}{\Gamma + i\omega}
\ee
provides us with a simple model for the dynamic response.
The dynamic susceptibility of a large spherical sample with parallel anisotropy axis and a simple cubic structure is shown in Fig. \ref{Fig: ac-interacting}.
In the overdamped case, the blocking temperature is not noticeably affected by 
the dipolar interaction while for low damping the blocking temperature decreases significantly as the interaction strength increases. 
These results are in agreement  with the simulations by Berkov and Gorn \cite{bergor2001} shown in Fig. \ref{Fig: berkov}.

An interpretation of the strong damping dependence found in the presence of a transverse field component was given in Ref. \cite{garsve2000} ---
the transverse field creates a saddle point in the potential barrier. A thermally excited spin  with high damping will ``fall'' directly back to the bottom of the potential well if the thermal excitation is not sufficient for an overbarrier jump. On the other hand, a weakly damped spin in the same situation will precess ($\sim 1/\lambda$ times) about the anisotropy axis and therefore has an increased probability for overbarrier jumps each time it passes close to the saddle point.
In the case of noninteracting particles the transverse field component must come from either a nonlinear probing field  \cite{garsve2000} or a bias field \cite{cofetal2001}, while the transverse field here naturally arises from the dipolar interaction.

\subsection{Numerical Methods}

Because of the long-range and reduced symmetry of the dipole-dipole interaction analytical methods such as the thermodynamic perturbation theory presented in Sec. \ref{Sec: TPT} will be applicable only for weak interaction. 
Numerical simulation techniques are therefore indispensable for the study of interacting nanoparticle systems, beyond the weak coupling regime.

The Monte Carlo (MC) method can be used to efficiently calculate thermal equilibrium properties (c.f. \fig{fig: Xeq-int}. However, since it is an energy-barrier-based method, it will fail to generate dynamic features such as the precession of the spins,
and will be able to generate the dynamic magnetization in the overdamped limit ($\lambda \to \infty$) only if an appropriate algorithm is used \cite{nowetal2000}.

Using a Langevin dynamics approach, the stochastic LLG equation [\eq{Eq: LLG}] can be integrated numerically, in the context of the Stratonovich stochastic calculus, by choosing an appropriate numerical integration scheme \cite{garpallaz98}.
This method was first applied to the dynamics of noninteracting particles \cite{garpallaz98} and later also to interacting particle systems \cite{bergor2001} (c.f. \fig{Fig: berkov}).

Because of the long-range nature of the dipolar interaction, care must be taken in the evaluation of the dipolar field. 
For finite systems the sums in \eq{Eq: Hlocal} are performed over all particles in the system. For systems with periodic boundary conditions the Ewald method \cite{ewa21,mad18,deleeetal80}, can be used to correctly calculate the conditionally convergent sum involved. However, in most work \cite{andetal97,bergor2001} the simpler Lorentz-cavity method is used instead.

\newcommand{\Tm}{T_{\rm m}}
\newcommand{\Tf}{T_{\rm f}}
\newcommand{\ising}{Fe$_{0.50}$Mn$_{0.50}$TiO$_3$ }
\newcommand{\AgMn}{Ag(11 at\% Mn){ }}
\newcommand{\agmn}{Ag(11 at\% Mn){ }}

\section{Strongly Interacting Nanoparticle Systems---\\Spin-Glass-Like Behavior}
\label{Chap: nanovxv}

A large number of applications use densely packed magnetic nanoparticles.
It is thus important to know how interparticle interaction affects the physical properties of magnetic nanoparticle systems.
In particular, it is important to understand how the thermal stability of magnetic recording media is changed by interparticle interactions due to the current effort of shrinking the volume of storage devices.

It has been suggested that dense nanoparticle samples may exhibit glassy dynamics due to dipolar interparticle interaction \cite{luoetal91}; disorder and frustration are induced by the randomness in the particle positions and anisotropy axis orientations.
In order to investigate  such systems, one needs to use the experimental techniques (protocols) developed in studies of spin glasses.
Examination of the effects of dipolar interactions using standard experimental protocols (zero-field-cooled ZFC / field-cooled FC magnetization) indicates no dramatic change in these quantities, and one can be misled to believe that the only effect of the dipolar interaction is to increase the blocking temperature due to enhanced energy barriers. However, glassy dynamics has been observed in strongly interacting nanoparticle systems \cite{jonetal95,mametal99,doretal99} and for strongly interacting systems with narrow size distributions evidence has been given for a spin-glass-like phase transition \cite{djuetal97,jonsvehan98,hanetal2002,sahetal2002}.

In frozen ferrofluids  the strength of the  dipolar interaction between the single-domain nanoparticles can be continuously varied by changing the particle concentration.
With increasing  particle concentration the magnetic behavior may evolve from superparamagnetic to spin-glass-like.
We will begin this section by recalling some fundamental properties of spin glasses. 
Furthermore, experimental results will be presented on a ferrofluid of Fe$_{1-x}$C$_x$ nanoparticles.
The glassy dynamics of dense nanoparticle samples will be compared with those of an Ising and a Heisenberg spin glass.

\subsection{Spin Glasses}

This section is intended as a brief introduction to spin glasses, focusing on the most recent work.
For reviews on spin glasses, see, for example, Refs. \cite{binyou86,fischerhertz,young}.

\subsubsection{Material}
The canonical spin glass consists of a noble metal (Au, Ag, Cu, or Pt) diluted with a transition metal ion, such as Fe or Mn. The magnetic interaction in such systems is mediated by the conduction electrons, leading to an indirect exchange interaction --- the RKKY (Ruderman and Kittel \cite{rudkit54}, Kasuya \cite{kas56}, and Yosida \cite{yos57}) interaction, whose coupling constant $J(R)$ oscillates strongly with distance $r$ between the spins as
\be
J(r) = J_0 \frac{\cos(2k_{\rm F} r + \varphi_0)}{(k_{\rm F} r)^3} \, .
\ee
Here $J_0$ and $\varphi_0$ are constants and $k_{\rm F}$ is the Fermi wavevector of the host metal.
Since the spins are randomly placed in the host metal, some spin--spin interaction will be positive and favor parallel alignment while other will be negative, thus favoring antiparallel alignment. 

The pure RKKY interaction is isotropic, and the canonical spin glass systems are therefore often referred to as {\em Heisenberg spin glasses}.
However, some anisotropy is also present in those systems originating from dipolar interaction  and interaction of the Dzyaloshinsky--Moriya (DM) type \cite{ferlev80}.
The latter is due to spin--orbit scattering of the conduction electrons by non-magnetic impurities and reads
\be
E_{\rm DM}=-\vec{D}_{ij} \cdot (\ei \times \ej ), 
\qquad \vec{D}_{ij} \propto \vec{r}_i \times \vec{r}_j \, ,
\ee
where $\vec{D}_{ij}$ is a random vector due to the randomness of the spin positions $\vec{r}_i$.
The dipolar interaction is, as discussed in Secion \ref{Sec: dipolar}, weak for atomic spin systems, while the DM interaction is enhanced by the presence of nonmagnetic transition-metal impurities \cite{ferlev80,prejol80}.
However, for macroscopic spins (magnetic moments), the dipolar interaction is important and if the particles are dispersed in a nonconductive medium (e.g., a frozen ferrofluid), it is the dominating interparticle interaction. If the magnetic particles are randomly placed, the dipolar interaction will be both positive and negative. 

A widely studied  model system for an Ising spin glass is single crystals of Fe$_x$Mn$_{1-x}$TiO$_3$ with $x \approx 0.5$ \cite{itoetal86,itoetal90,aruito93}. 
Both FeTiO$_3$ and MnTiO$_3$ are antiferromagnets having the easy axis along the hexagonal $c$ axis of the ilmenite structure.  
The Fe$^{2+}$ spins in  FeTiO$_3$ are coupled ferromagnetically within a $c$ layer and antiferromagnetically between adjacent $c$ layers. In MnTiO$_3$, on the other hand, both the intralayer and interlayer coupling of Mn$^{2+}$ spins are antiferromagnetic.
The compound Fe$_x$Mn$_{1-x}$TiO$_3$ behaves as an Ising spin glass for $0.4 \lesssim x \lesssim 0.57$, due to the mixing of ferromagnetic and antiferromagnetic interaction \cite{aruito93}.

There are essential differences between the systems we have presented here, even in the limit of very strong anisotropy. 
The RKKY spin glasses are Heisenberg systems with {\em random  unidirectional} anisotropy. 
Ferrofluids frozen under zero field are Heisenberg systems with {\em random  uniaxial} anisotropy, while an Ising system is characterized by {\em parallel uniaxial} anisotropy.

\subsubsection{Spin Glass Models}

The Hamiltonian of an Ising spin glass, given by Edwards and Anderson (EA) \cite{edwand75}, is 
\be
\Hamil = - \case{1}{2} \sum_{i,j} J_{ij} s_i s_j 
- 
H \sum_{i=1}^N s_i\, , 
\label{Eq: H-sg}
\ee
where the spin $s_i=\pm 1$ and the coupling constants $J_{ij}$ are chosen from some random distribution fulfilling  $\sum_{i,j}J_{ij}=0$ in the case of a symmetric spin glass. In the EA model the spin--spin interaction is only of the nearest-neighbor type.
The Sherrington--Kirkpatrick (SK) model \cite{shekir75} is the infinite-range version of the EA model.
It is most useful as a basis for mean-field calculations. One such solution is the replica symmetry breaking theory of Parisi \cite{par79,par80,par83}.

For a ferromagnet the order parameter is the magnetization, while for antiferromagnets it is the sublattice magnetization.
For spin glasses the magnetization is zero at all temperatures, and an appropriate order parameter was proposed by Edwards and Anderson \cite{edwand75} as the average value of the autocorrelation function
\be
q_{\rm EA} = \lim_{t \to \infty} \la \ei(0) \cdot \ei(t) \ra \, .
\ee 
The order parameter susceptibility, which diverges at the transition temperature, is the nonlinear susceptibility $\chi_2$, defined as $M/h =\chi_0 +\chi_2 h^2 + \cdots$, where $\chi_0$ is the linear susceptibility \cite{suzuki77}.
The divergency of the nonlinear susceptibility was first shown on a Au(Fe) spin glass in 1977 by Chikazawa et al. \cite{chietal79} and more recently for a strongly interacting nanoparticle system by Jonsson et al. \cite{jonsvehan98}.

\subsubsection{Critical Dynamics}

Close to the transition temperature $\Tg$,  the dynamics of a spin glass system will be governed by critical fluctuations, but critical fluctuations are also of importance on experimental timescales quite far from $\Tg$. At temperatures both below and above $\Tg$, length scales shorter than the coherence length of the critical fluctuations 
\be
\xi \sim L_0 |\epsilon|^{-\nu} \, ,
\ee
will  be dominated by critical fluctuations.
Here $L_0$ is a microscopic length scale and the reduced temperature $\epsilon=(1-T/\Tg)$.
The coherence  length  can be transformed into a timescale according to conventional critical slowing down; the critical correlation time is given by
\be
\tau_c \sim \tau_m (\xi(T)/L_0)^z \sim \tau_m |\epsilon|^{-z\nu} \, ,
\label{Eq: tauc}
\ee
where $\tau_m$ is a microscopic timescale.
For spin glasses,
$\tau_m \sim 10^{-13}$ s is the fluctuation time of an atomic
moment. For nanoparticles, $\tau_m$ can be assigned to the
superparamagnetic relaxation time of a single particle of average
size, which, in the relevant temperature range for our studies,
can be approximated by the Arrhenius--N{\'e}el expression [\eq{Eq: arrhenius}].

\begin{figure}[t]
\begin{center}
\includegraphics[width=0.75\textwidth]{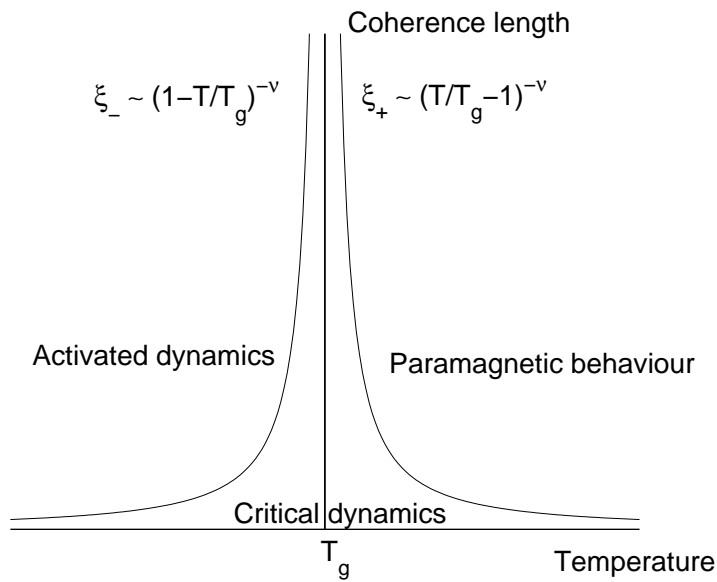}
\caption{Illustration of length (time) scales in spin glasses.
\label{Fig: length-scales}}
\end{center}
\end{figure}

At temperatures below $\Tg$, there is a crossover between critical dynamics on short length (time) scales  and activated dynamics on long length (time) scales. The length scale of critical dynamics as a function of temperature is illustrated in Fig. \ref{Fig: length-scales}.
For $T>\Tg$, the system is in equilibrium on length (time) scales longer than $\xi$ ($\tau_c$) and hence the magnetic response is paramagnetic.
The ``freezing'' temperature ($T_{\rm f}(\tau_c)$) of the crossover from a paramagnetic response to slow spin glass dynamics, as a function of observation time, can be obtained from experiments. 
Such a dynamic scaling analysis is performed in Sec.~\ref{sec-trans} for two samples of a nanoparticle system with two different volume concentrations.
Dynamic scaling analyzes have given evidence for critical slowing down in a wide range of spin glass materials \cite{binyou86,fischerhertz,norsve97}.
The critical exponents obtained are rather scattered. However, the critical exponents of Ising and Heisenberg systems are clearly different \cite{petit-thesis,jonetal2002PRL}.

\subsubsection{Nonequilibrium Dynamics}

\begin{figure}[t]
\center
\includegraphics[width=0.75\textwidth]{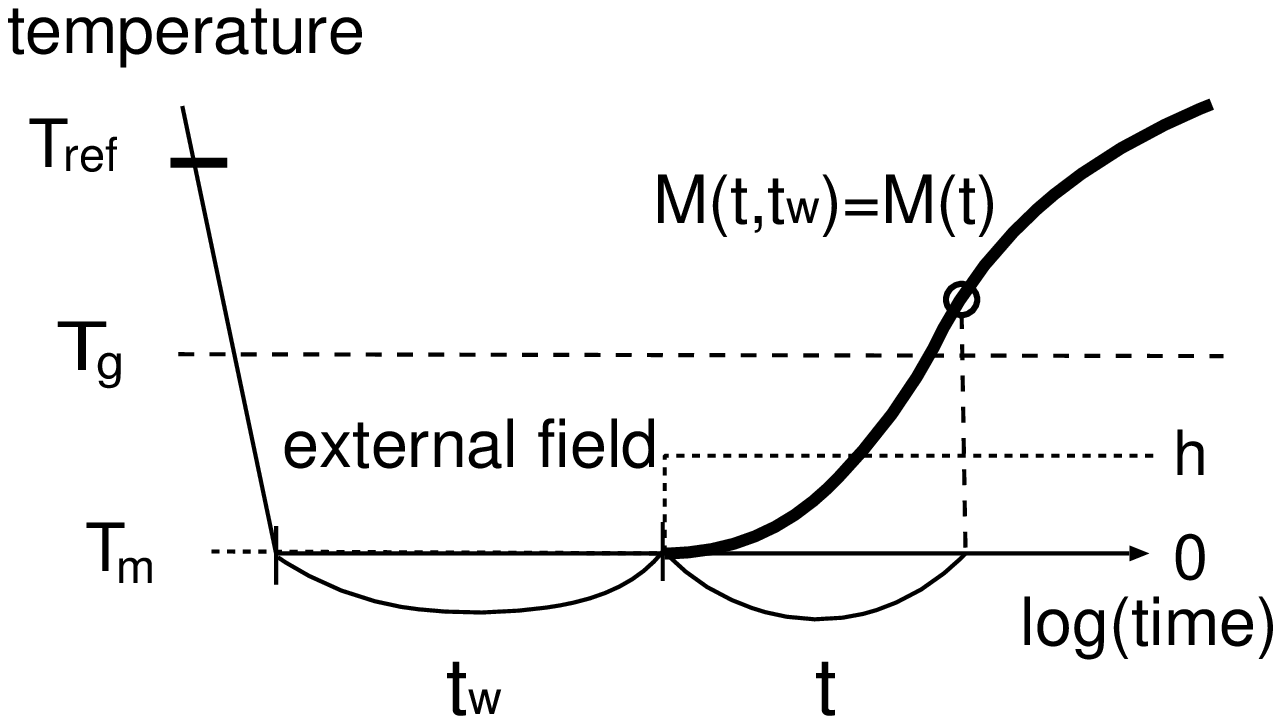}
\caption{Schematic representation of a ZFC relaxation experiment.
The sample is cooled to the measurement temperature $\Tm$ under zero field, after a wait time $\tw$, a small probing field $h$ is applied and the ZFC magnetization is recorded as a function of time.
\label{Fig: relax}}
\end{figure}

\begin{figure}[hbt]
\begin{center}
\includegraphics[width=0.75\textwidth]{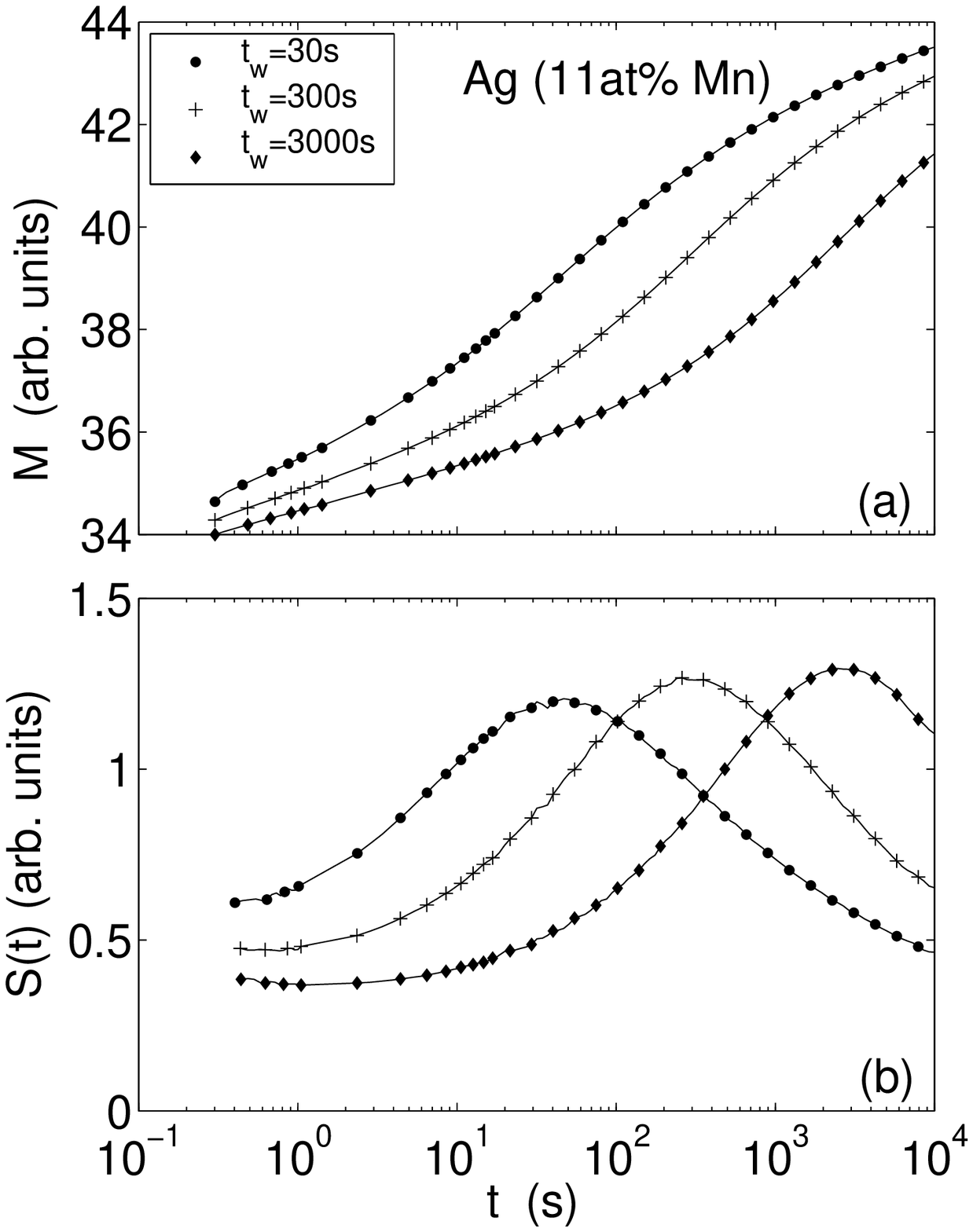}
\caption{(a) ZFC susceptibility versus time on a logarithmic scale for the Ag(11 at\% Mn) spin glass. $\Tm=30$~K and $h=0.5$~G.
(b) Relaxation rate $S(t) = h^{-1} \frac{\partial M}{\partial \log t}$ of the  ZFC susceptibility curves shown in (a).
\label{Fig: AgMn_S2}}
\end{center}
\end{figure}

Aging phenomena in glassy materials were first discovered and thoroughly investigated in the field of structural glasses \cite{struik78}. Magnetic aging in spin glasses was first observed by Lundgren et al. in 1983 \cite{lunetal83}.
It was found that the ZFC relaxation depends on the wait time $\tw$ during which the system has  been allowed to age at the measurement temperature, before applying the  probing field and recording the magnetization as a function of time $t$. The measurement protocol is illustrated in \fig{Fig: relax} and ZFC relaxation measurements on a Ag(11 at\% Mn) spin glass are shown in  Fig.~\ref{Fig: AgMn_S2}.
It can be seen that the relaxation rate $S(t) = h^{-1} dM / d\log t$ exhibits a maximum at $t \sim \tw$ \cite{lunetal83,andmatsve92}.
Many different types of materials were later shown to exhibit aging and nonequilibrium dynamics, including  polymers \cite{belcillar2000}, orientational glasses \cite{douetal99}, gels \cite{noretal2000}, and ceramic superconductors \cite{papetal99}.


Spin glass systems represent ideal model systems for studying nonequilibrium dynamics experimentally, numerically, and theoretically.
The dynamic properties of spin glasses can accurately be  investigated by {\it superconductive quantum interference device} (SQUID) magnetometry on specially designed magnetometers with low-background field ($<1$~mOe) and optimized temperature control ($\Delta T <100 \,{\rm \mu K}$) \cite{magetal97}.
Spin glasses exhibit nonequilibrium dynamics on all timescales from $\tau_m$ to infinity. 
Experimental studies are, however, limited to a finite time window $t/\tau_m \sim 10^{8} - 10^{17}$, while numerical simulations are limited to much shorter timescales  $t/\tau_m \sim 1 - 10^{5}$. 
Because of the larger spin--flip time of a magnetic moment compared to an atomic spin, interacting nanoparticle systems can be investigated in an experimental time window in between those of ordinary spin glasses and numerical simulations. 
In addition, the strength of the dipolar interaction can be tuned by the particle concentration of the nanoparticle system.
Strongly interacting nanoparticle systems are therefore also interesting as model systems for glassy dynamics.

\paragraph{The Droplet Model}

The ``droplet'' theory \cite{mcm84,bramor86,fishus86,fishus88eq,fishus88noneq} is a real-space theory, based on renormalization group arguments for the Ising EA model with a continuous distribution of independent exchange.
It makes predictions concerning the nonequilibrium dynamics within the spin glass phase.
Important concepts are domain growth --- growth of the coherence length for equilibrium spin-glass order, and temperature chaos.
Both these concepts can also be applied to less anisotropic spin glasses \cite{jonetal2002PRL,bouetal2001,dupetal2001,jonyosnor2002} and generalized to other glassy systems \cite{beretal}.

At each temperature the equilibrium spin glass state is considered to consist of a ground state plus thermally activated droplet excitations of various sizes. 
A droplet is a low-energy cluster of spins with a volume $L^d$ and a fractal surface area $L^{d_s}$.
The typical droplet free-energy scales as
\be
F_L^{\rm typ} \sim \Upsilon(T) \left(\frac{L}{L_0}\right)^{\theta} \, , 
\qquad \Upsilon(T) \sim J \epsilon^{\theta \nu} \, ,
\ee
where $\theta$ is the stiffness exponent and $\Upsilon(T)$ is the stiffness.
The droplet free energy is broadly distributed, and because of the presence of configurations that are almost degenerate with the ground state, the distribution of $F_L$ will have weight down to zero energy:
\be 
\rho_L(F_L) \approx \frac{\tilde{\rho}(F_L/F_L^{\rm typ})}{F_L^{\rm typ}}, 
\qquad \tilde{\rho}(0) >0.
\ee
If $\theta < 0$, large droplets can be flipped at a low energy cost so that the large droplets will not be stable against small fluctuations and the system will be paramagnetic.  
Hence, a negative value of $\theta$ indicates that the system is below its lower critical dimension \cite{mcm84,bramoo84}. 
On the other hand, if $\theta >0$, very few of the large-scale droplets will be thermally activated since $F_L^{\rm typ} > \kT$. Since $\rho_L(F_L)$ has non-zero weight near zero energy, a certain fraction of droplets will be thermally active and dominate most of the equilibrium physics.

The dynamics of droplets is considered to be a thermally activated process. The energy barrier for annihilation of a droplet will scale as
\be
B_L^{\rm typ} \sim \Delta(T) \left(\frac{L}{L_0}\right)^{\psi} \, , 
\qquad \Delta(T) \sim J \epsilon^{\psi \nu} \, ,
\ee
where $\Delta(T)$ sets the free-energy scale of the barriers and $\psi$ is an exponent satisfying $\theta < \psi < d-1$.
The characteristic time $\tau_L$ that a thermally activated droplet will last for is given by an Arrhenius law
\be
\ln \left[\frac{\tau_L}{\tau_0(T)}\right] \sim \frac{B_L}{\kT} \, ,
\ee
where $\tau_0(T)$ is the unit timescale for the activated process. 
For activated hopping processes the unit timescale is {\em not} simply given by the real microscopic timescale \cite{risken}, which is $\tau_m \sim \hbar/J \sim 10^{-13}$ s in spin systems. 
A plausible choice for $\tau_{0}(T)$ is instead the critical correlation time 
$\tau_c$ [\eq{Eq: tauc}] as proposed in Ref. \cite{fishus88noneq}.
The Arrhenius law implies that droplets of length scale $L=L_T(t)$
\begin{equation}
L_T(t) 
\sim \left[ \frac{\kT \ln (t/\tau_{0}(T))}{\Delta(T)} \right]^{1/\psi} \,,
\label{Eq: L-log}
\end{equation}
can be activated within a timescale $t$.
 
$L_T(\tw)$ will be the characteristic length scale of equilibrium spin glass order at a time $\tw$ after  a quench from a temperature above $\Tg$ to a temperature $T$ in the spin glass phase.

In a magnetization measurement the system is probed by applying a small magnetic field. 
The magnetization arises through the polarization of droplets. 
Since this polarization also is a thermally activated process, it will affect droplets of size $L(t)$, where $t$ is the time elapsed after the application of the magnetic field in a ZFC-relaxation experiment or $t=1/\omega$ in an ac experiment at a given angular frequency $\omega$.
The peak observed in the relaxation rate of the ZFC magnetization at $t \approx \tw$ (see Fig.~\ref{Fig: AgMn_S2}) has, within the droplet model, been interpreted as a crossover from quasiequilibrium dynamics at $L(t) < L(\tw)$ to  nonequilibrium dynamics at $L(t) > L(\tw)$ (see Ref.~\cite{yoshuktak2002} for a detailed discussion).

According to the droplet theory, typical spin configurations
of a pair of equilibrium states at two different temperatures, 
say, $T_{1}$ and 
$T_{2}$, are essentially the same on short 
length scales much below the so-called overlap length $\Lovlp$, but 
completely different on large length scales much beyond $\Lovlp$.  
This temperature chaos is due to a subtle competition between energy and entropy in the spin glass phase.
In the limit of small temperature differences $|\Delta T/J| \ll 1$,
the overlap length between the two temperatures $T_{1}$
and  $T_{2}=T_{1} + \Delta T$ is supposed to scale as \cite{fishus88eq,fishus88noneq,bramoo87}
\be
\Lovlp \sim L_{0} \left| \frac{\Delta T}{J}\right|^{-\zeta} \, , 
\qquad 
\zeta=\frac{2}{(d_{\rm s}-2\theta)} \, ,
\label{Eq: lovlp}
\ee
where $\zeta$ is the chaos exponent.

Experimentally, temperature chaos can be evidenced by aging the system at a temperature $T_i$ and changing the temperature to a $T_i +\Delta T$. If the temperature shift is large enough, the equilibrium domain configurations at $T_i$ and $T_i +\Delta T$ are completely different on the length scales relevant for the experiment. Hence, if the magnetic response at  $T_i +\Delta T$ is the same as after a direct quench, the system appears {\em rejuvenated}.

\paragraph{Experiments: Aging, Memory and Rejuvenation}
\label{sec-noneq-sg}
After discovery of the aging effect in the spin glass phase, experimental protocols with temperature steps and cyclings were proposed \cite{graetal88}.  
These experiments showed not only rejuvenation effects but also that spin glass order, characteristic of different temperatures, can coexist on different length  scales; hence the spin glass keeps a {\em memory} of its thermal history.

\begin{figure}[htb]
\center
\includegraphics[width=0.6\textwidth]{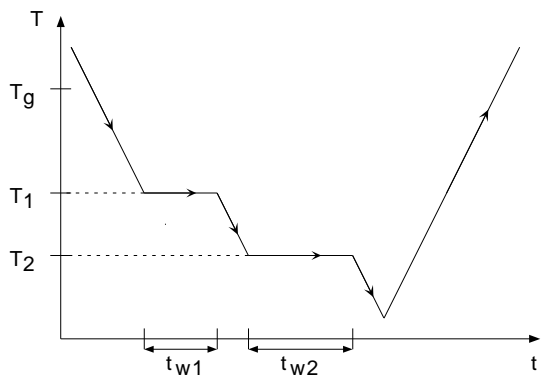}
\caption{
The experimental procedure of a ``memory'' experiment with two halts on cooling.
\label{Fig: memory}}
\end{figure}

\begin{figure}[thb]
\center
\includegraphics[width=0.75\textwidth]{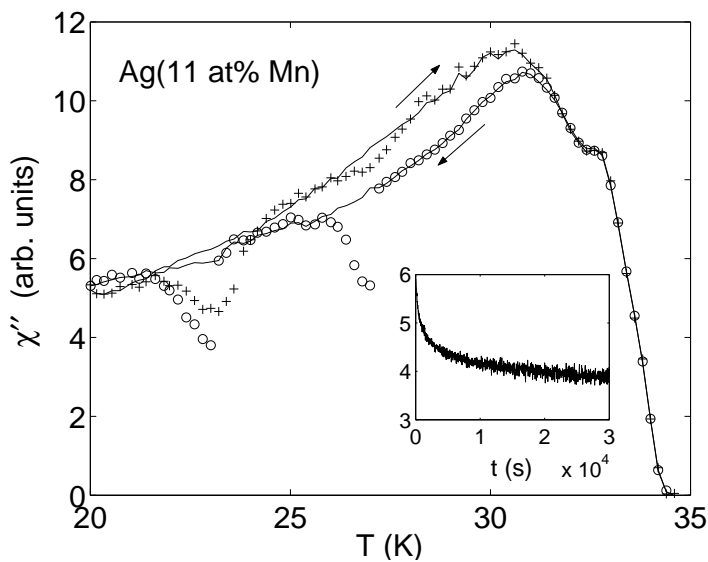}
\caption{
$\chi''(\omega) $ versus temperature for the \AgMn sample measured during cooling (circles) and during the subsequent reheating (pluses). Two intermittent halts were made during the cooling: one at 27 K for $10,\,000$ s and another at 23~K for $30,\,000$~s. The susceptibility measured on constant cooling and on constant heating is shown as reference (solid lines). The arrows indicate the cooling and heating curves, respectively.
The inset shows $\chi''(\omega) $ versus time during  the halt at $T=23$~K; $\omega/2\pi = 510$ mHz. The cooling and heating rate is $\sim 0.2$~K/min.
\label{Fig: AgMn-ac-mem}}
\end{figure}

\begin{figure}[thb]
\center
\includegraphics[width=0.75\textwidth]{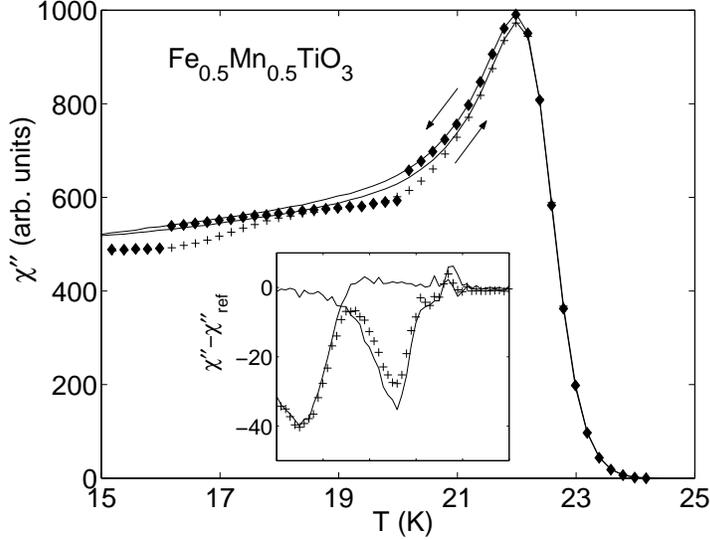}
\caption{
a) $\chi''$ versus temperature measured on cooling (diamonds), with two intermittent stops at 20 K for 10,\,000 s and at  16 K for 30,\,000~s, and on the subsequent reheating (pluses), for the \ising sample.
The reference cooling and heating curves  are drawn with solid lines.
Inset: $\chi''-\chi''_{\rm ref}$ versus temperature derived from the  heating curves for single and double stops. $\omega/2 \pi=510$~mHz. 
\label{Fig: Ising-ac-mem}}
\end{figure}

A simple experimental protocol \cite{norsve97} was employed to illustrate memory and rejuvenation effects \cite{jonetal98,jonetal99,matetal2001,matetal2002}; 
the sample is cooled from a high temperature with one (or more) halts of the cooling at one (or more) temperatures in the spin glass phase.
The experimental procedure of a double-stop experiment is illustrated in Fig.~\ref{Fig: memory}.
The ac susceptibility is subsequently recorded on heating. 
An ``ac memory'' experiment is shown in Fig.~\ref{Fig: AgMn-ac-mem} for a \agmn spin glass and in Fig.~\ref{Fig: Ising-ac-mem} for an \ising spin glass.
The ac susceptibility is measured on cooling with two intermittent stops.
During a stop the ac susceptibility relaxes downward as shown in the inset of Fig.~\ref{Fig: AgMn-ac-mem}.
The level of the ac susceptibility is hence related to the age of the system (a lower susceptibility indicates an older system).
As the cooling is resumed, the ac susceptibility merges (quite rapidly) with the reference curve, and the system is {\em rejuvenated}. 
On subsequent reheating the ac susceptibility shows a dip around each  aging temperature --- the system has kept a {\em memory} of each equilibration at constant temperature.
The memory experiment is an efficient tool to study spin-glass-like properties in various materials \cite{jonhannor2000,coletal2000,belcillar2002,kitetal2002,garetal2003}.

For the \AgMn sample, the ac susceptibility curve measured on cooling lies below the curve subsequently measured on heating, except close to the lowest temperature (see Fig. \ref{Fig: AgMn-ac-mem}).
If the aging at different temperatures is accumulative, the heating curve would appear older than the cooling curve and therefore lower in amplitude.
For the \ising sample this is indeed the case (see Fig. \ref{Fig: Ising-ac-mem}).
The nonaccumulative behavior observed in the cooling--heating curves of the ac susceptibility of the \AgMn sample can qualitatively be explained by rejuvenation during cooling and heating due to strong temperature chaos. 
For the \ising sample, the double-memory experiment shown in Fig. \ref{Fig: Ising-ac-mem} indicates that temperature chaos does exist, but that it is much weaker than for \AgMn.

The memory experiments presented above yield, in a simple and illustrative way, information about aging, memory, and rejuvenation effects for the two spin glass systems.
However, cooling and heating rate effects as well as memory and rejuvenation phenomena are all mixed in a nontrivial way.
Specially designed thermal protocols have been used to quantitatively investigate aging and rejuvenation proprieties of the \ising and \agmn sample \cite{dupetal2001,jonetal2002PRL,jonyosnor2002,jonetal2003}.
It was shown that the temperature dependence of $L_T(t)$ is stronger for the \agmn sample than for the \ising sample. Also, the rejuvenation effects are much stronger for the \agmn sample.


\subsection{FeC Nanoparticle Systems}

\begin{figure}[htb]
\center
\includegraphics[width=0.9\textwidth]{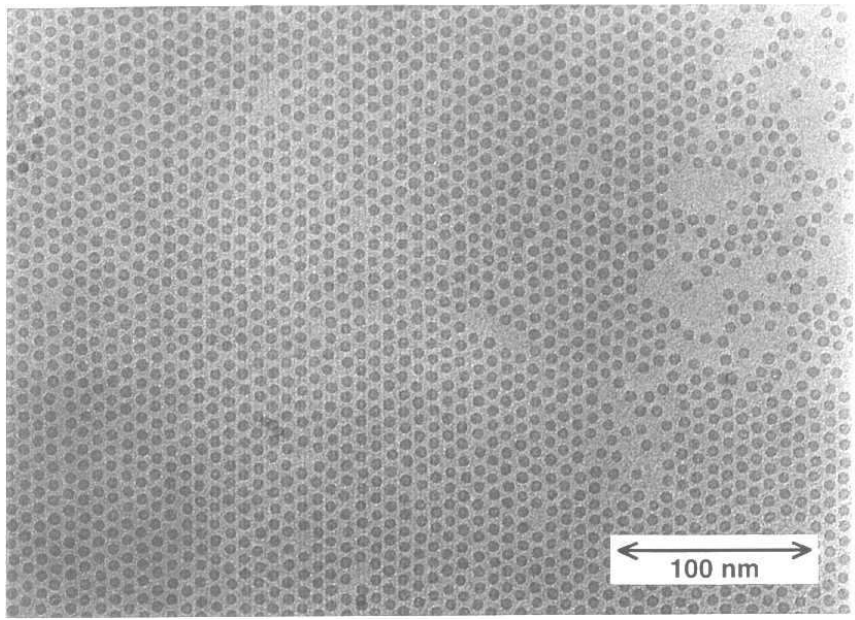}
\caption{TEM picture of typical Fe$_{1-x}$C$_{x}$ nanoparticles.
\label{Fig: TEM}}
\end{figure}

We now focus on  a ferrofluid of single-domain particles of the amorphous alloy Fe$_{1-x}$C$_{x}$ ($x$ $\approx$ 0.2--0.3).
The particles were coated with a surfactant (oleic acid) and dispersed in a carrier liquid (xylene).
The particle shape is nearly spherical (see Fig \ref{Fig: TEM}) and the average particle diameter $d =5.3 \pm 0.3$~nm.
The saturation magnetization was estimated to $\Ms = 1\times 10^6$ A/m, the microscopic flip time to $\tau_0 = 1\times 10^{-12}$~s, and  the uniaxial anisotropy constant $K = 0.9\times 10^5$ J/m$^{3}$ \cite{hanetal2002}.
The interparticle interaction can be varied by changing the particle concentration of the ferrofluid. 
The strength of the interaction for a given concentration is determined by the anisotropy constant and the saturation magnetization according to \eq{Eq: hd}
with the parameters given above  $\hd \approx 0.56 c$, where $c$ is the volume concentration of nanoparticles.
The samples studied here originate from the same batch as those in Refs. \cite{jonhannor2000,jonetal2001JMMM,jonetal2001PRB,hanetal2002}. Earlier studies use samples from different batches having slightly different physical properties \cite{hanetal95a,hanetal95b,djuetal97,jonsvehan98,hanetal98}.

\begin{figure}[th]
\center
\includegraphics[width=0.75\textwidth]{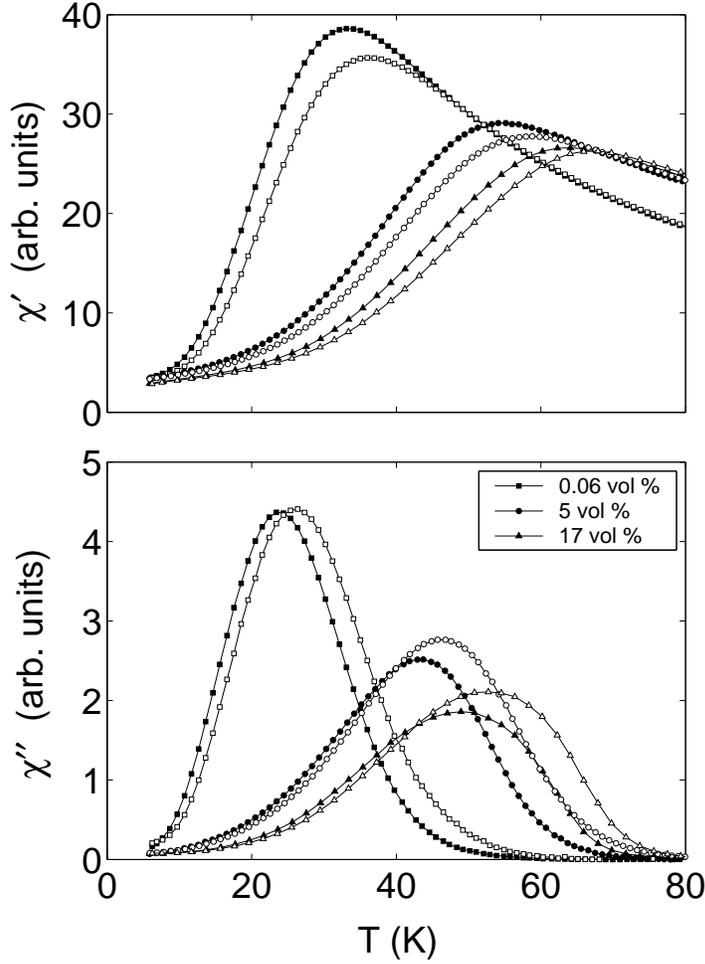}
\caption{Ac susceptibility vs temperature at frequencies $\omega/2\pi=125$ Hz (filled symbols) and $\omega/2\pi=1000$ Hz (open symbols).
\label{Fig: ac-susc}}
\end{figure}

Figure \ref{Fig: ac-susc} shows the real and imaginary part of ac susceptibility versus temperature for three different particle concentrations of the FeC sample: $c=0.06$, 5, and 17 vol\%. With increasing concentration, the peak in the ac susceptibility is shifted to higher temperatures and the curve is simultaneously  suppressed. 
This behavior is different from that of the weakly interacting nanoparticle systems shown in Figs. \ref{Fig: berkov} and \ref{Fig: ac-interacting}.
In this section we will argue that the dynamics of the 5 and 17 vol\% samples is spin-glass-like, and hence fundamentally different from the superparamagnetic behavior of noninteracting and weakly interacting nanoparticle systems. 

\subsubsection{Nonequilibrium Dynamics}

\begin{figure}[t]
\center
\includegraphics[width=0.75\textwidth]{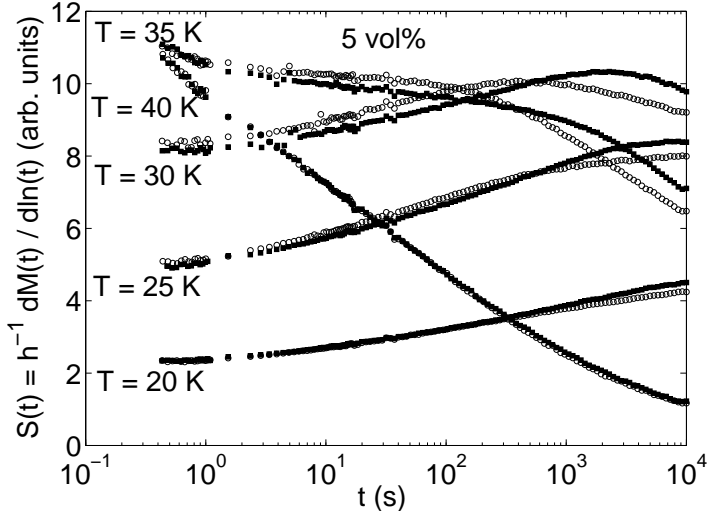}
\caption{$S(t)$ versus time on a logarithmic scale for the 5 vol\% sample obtained from ZFC relaxation measurements with $\tw= 300$~s (open symbols) and 3000~s (filled symbols); $h=0.05$~Oe.
\label{Fig: aging}}
\end{figure}

Magnetic aging can be evidenced by measuring the ZFC relaxation at constant temperature after a fast cooling through the transition temperature using different wait times before applying the magnetic field.  The experimental procedure was depicted in Fig.~\ref{Fig: relax} together with a typical measurement on a \agmn  spin glass in Fig.~\ref{Fig: AgMn_S2}.
In a noninteracting nanoparticle system, the low-field ZFC relaxation is governed only by the distribution of relaxation times of the particles and their temperature dependence. It does not depend on the wait time at $\Tm$ before applying the probing field,\footnote{The only wait time dependence that could exist is the adjustment of the position of the magnetic moment to the Boltzmann distribution at $\Tm$. This adjustment does however only involve intrawell rotation and is therefore a much faster process than the slow cooling to $\Tm$ and the time needed to stabilize the temperature.} as was shown experimentally in Ref.~\cite{jonetal95}. 
The relaxation rate $S(t)$ of the ZFC magnetization measured
 at different temperatures between 20 and 40~K are shown in  Fig. \ref{Fig: aging} for the 5 vol\% sample. 
The measurements are repeated for two different wait times 300 and 3000~s.
A clear difference between the $S(t,\tw)$ curves for $\tw=300$ and 3000~s can be seen for all temperatures $< 40$~K, presenting evidence for glassy nonequilibrium dynamics at those temperatures.
The shapes of the $S(T)$ curves are, however, rather different from those of canonical spin glasses (see Fig.~\ref{Fig: AgMn_S2}).
A peak in $S(t)$ at $t \sim \tw$ as in ordinary spin glasses is observed only at $T=30$~K. The difference arises largely from the strong temperature dependence of the individual particle relaxation time compared to the almost temperature independent relaxation time of individual spins.

\begin{figure}[ht]
\includegraphics[width=0.75\textwidth]{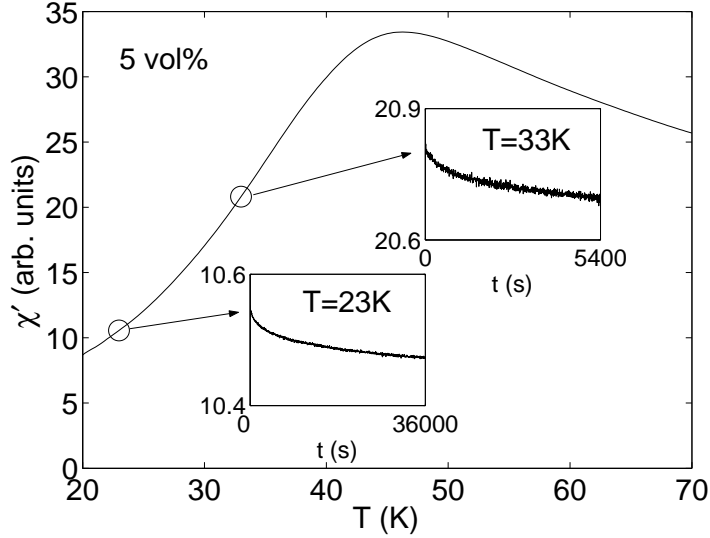}
\center
\caption{$\chi'$ versus temperature for the 5 vol\% sample. The insets show how $\chi'$ relaxes with time if the cooling is halted at 33 or 23~K.  
$f=510$~mHz
\label{Fig: acT_time}}
\end{figure}

The 5 vol\% sample has also been investigated with ac susceptibility measurements. $\chi'(T)$ is shown in Fig.~\ref{Fig: acT_time} for a low frequency. The insets show how the ac susceptibility relaxes with time if the cooling is halted at $T_s=33$ or 23~K. 
The relative relaxation of the ac susceptibility is smaller than for ordinary spin glasses. $\chi'$ relaxes more in absolute units than $\chi''$ and $\chi'$ has therefore been chosen to illustrate the nonequilibrium dynamics.
For a waiting time of 30,\,000~s at low temperature $\Delta \chi ''/\chi_{\rm ref}'' \approx 3$ \%  compared to $\approx 8$ \% for the \ising sample (see Fig.~\ref{Fig: Ising-ac-mem}) and $\approx 30$ \% for the \agmn sample (see Fig.~\ref{Fig: AgMn-ac-mem}).

\begin{figure}[t!]
\center
\includegraphics[width=0.75\textwidth]{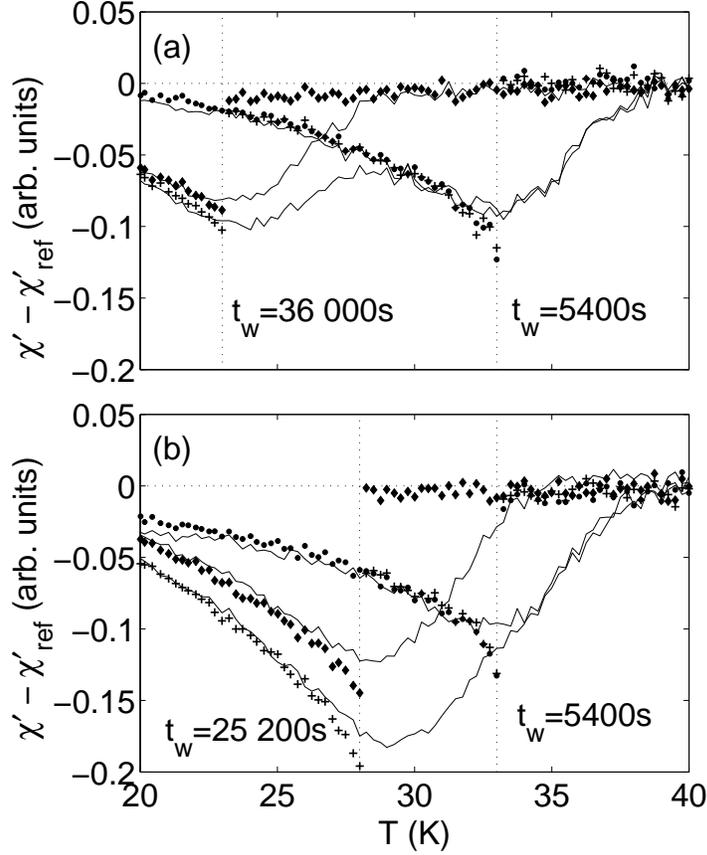}
\caption{$\chi'(T)-\chi'_{\rm ref}(T)$ versus $T$ measured on
cooling (symbols) and heating (lines) for the 5 vol\% sample. 
Circles---the cooling was halted at 33~K for 5400~s; 
diamonds---the cooling was halted at
(a) 23~K for 36\, 000~s, (b) 28~K for 25,\,200~s; pluses---the cooling was halted at $T_1 = 33$~K
for $t_{\rm w_{1}} = 5400$~s and at (a) $T_2 = 23$~K for $t_{\rm w_{2}} = 36\,000$~s, (b) $T_2 = 28$~K for $t_{\rm w_{2}} = 25\,200$~s. $\chi_{\rm ref}'$ is shown in Fig.~\ref{Fig: acT_time}.
$\omega/2\pi=510$ mHz.
\label{Fig: FeC-ac-mem}}
\end{figure}

Memory experiments with one and two temporary stops during cooling are shown in  Fig.~\ref{Fig: FeC-ac-mem}.
The nonequilibrium effects are more clearly revealed by subtracting the reference curves obtained on constant cooling and reheating.
The features are qualitatively similar to those of ordinary spin glasses (see Sec. \ref{sec-noneq-sg});
during a halt in the cooling, the system ages --- the ac susceptibility decreases.
When the cooling is resumed, the ac susceptibility slowly regains the reference level.
On the subsequent heating, the ac susceptibility exhibits a ``dip''centered around $T_{\rm s}$ in a single-stop experiment. In a double-stop experiment it exhibits two dips if the two halts are well separated in temperature [Fig.~\ref{Fig: FeC-ac-mem}(a)], but if the two halts are close in temperature [Fig.~\ref{Fig: FeC-ac-mem}(b)], only one large dip is observed on heating as a result of the two aging processes. 
In both cases, the difference curve $\chi'-\chi_{\rm ref}'$  of the double-stop experiment equals the sum of $\chi'-\chi_{\rm ref}'$ of the two single-stop experiments.  
As for the \ising sample, the ac susceptibility measured on heating lies below the one measured on cooling. 
In conclusion,  the rejuvenation effect in particle systems is weaker than in ordinary spin glasses and as a consequence, the memory of both one and two aging processes is better preserved on reheating than in the spin glass case.

This nanoparticle sample exhibits strong anisotropy, due to the uniaxial anisotropy of the individual particles and the anisotropic dipolar interaction.
The relative timescales ($t/\tau_m$) of the experiments on nanoparticle systems are shorter than for conventional spin glasses, due to the larger microscopic flip time.
The nonequilibrium phenomena observed here are indeed rather similar to those observed in numerical simulations on the Ising EA model \cite{komyostak2000A,picricrit2001}, which are made on  much shorter time (length) scales than experiments on ordinary spin glasses \cite{berbou2002}.

\subsubsection{A Spin Glass Phase Transition?}
\label{sec-trans}
Aging and nonequilibrium dynamics indicate but give by no means evidence for a thermodynamic phase transition at finite temperature to a low-temperature spin-glass-like phase. 
For example, two-dimensional (2D) spin glasses ($\Tg=0$) have been shown to exhibit similar nonequilibrium dynamics as 3D spin glasses \cite{matetal93,schetal93prb}.
A detailed analysis of the magnetic response close to the assumed transition temperature is needed in order to evidence a spin glass phase transition (see, e.g., Ref. \cite{norsve97} and references cited therein).

The existence of a second-order phase transition can be evidenced from critical slowing down [\eq{Eq: tauc}] approaching the phase transition from the paramagnetic phase. 
Defining a criterion to determine the freezing temperature $T_{\rm f}$ associated with a certain relaxation time, it is possible to derive the ``transition'' line ($\tau_c$) between thermodynamic equilibrium and critical dynamics as in Fig. \ref{Fig: length-scales}. 
In ac susceptibility measurements the relaxation time $\tau = 1 / \omega$, and a possible criterion for the freezing is where $\chi''(T,\omega)$ attains a certain fraction, say, 15\%, of its maximum value.

\begin{figure}[th]
\center
\includegraphics[width=0.8\textwidth]{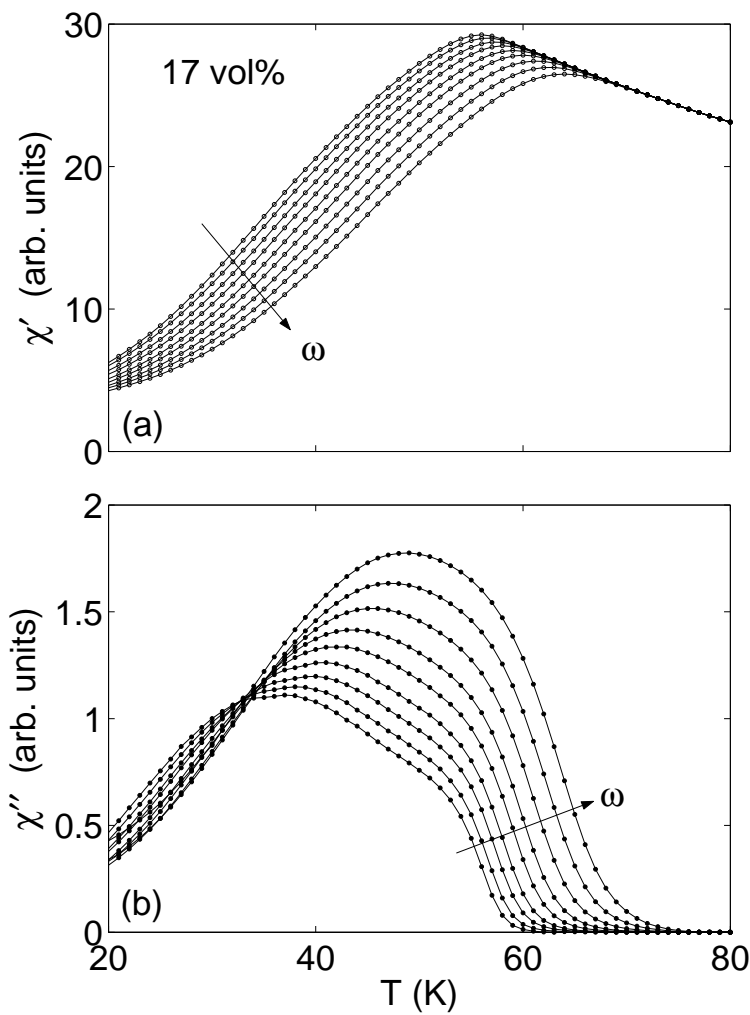}
\caption{ac susceptibility versus temperature for the 17 vol\% sample.
$\omega/2\pi= 0.017$, 0.051, 0.17, 0.51, 1.7, 5.1, 17, 55, 170~Hz.
\label{Fig: ac17}}
\end{figure}

\begin{figure}[h!]
\center
\includegraphics[width=0.8\textwidth]{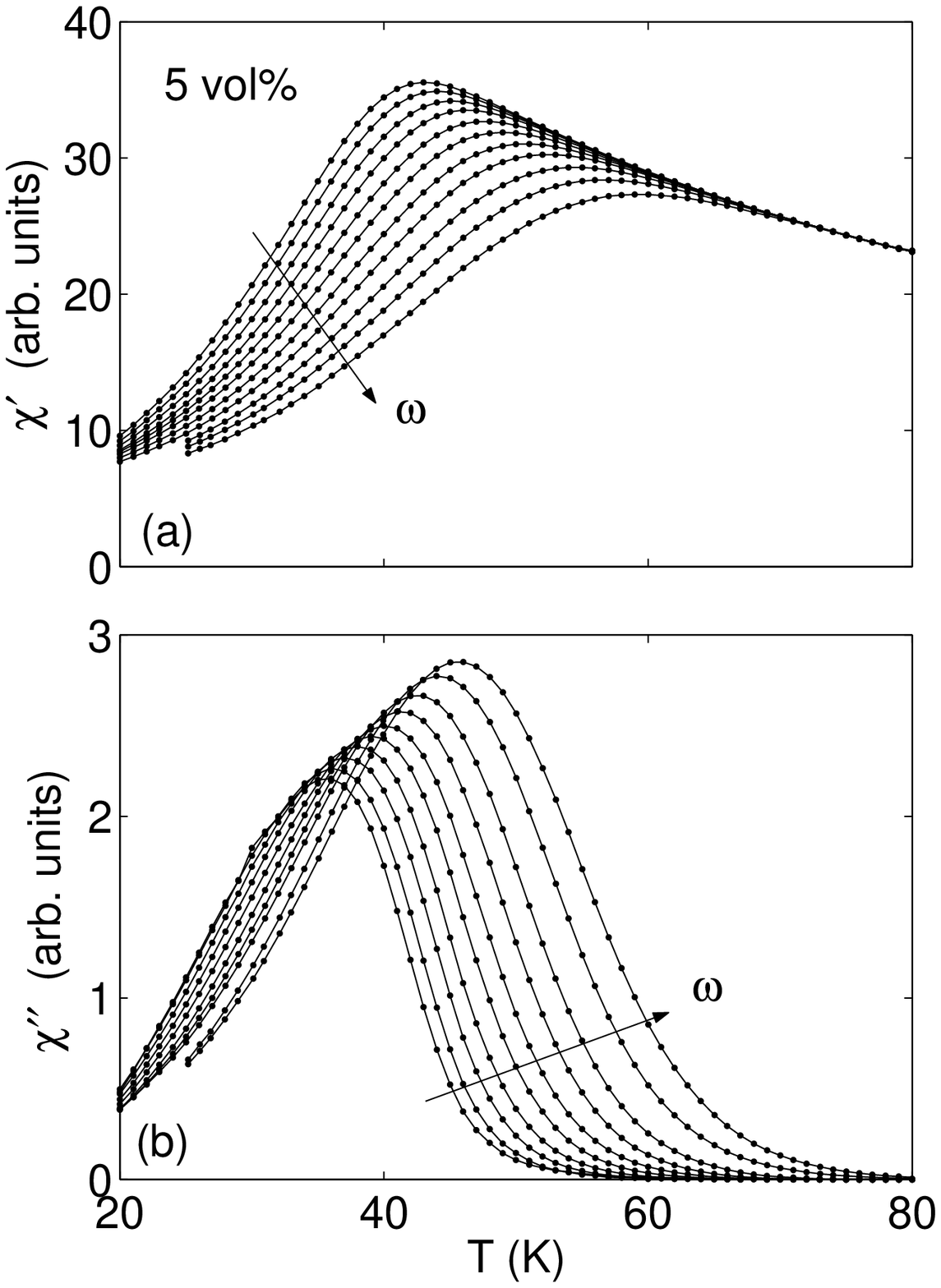}
\caption{ac susceptibility versus temperature for the 5 vol\% sample.
$\omega/2\pi= 0.017$, 0.051, 0.17, 0.51, 1.7, 5.1, 17, 55, 170, 510, 1700~Hz.
\label{Fig: ac5}}
\end{figure}

\begin{figure}[htb]
\center
\includegraphics[width=0.8\textwidth]{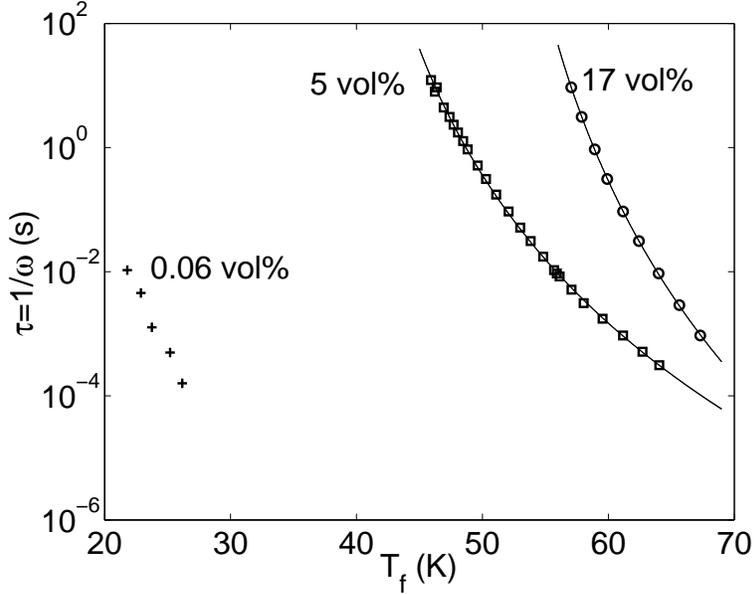}
\caption{Relaxation time $\tau = \omega^{-1}$ versus $T_{\rm f}$. For the 5 vol\% and 17 vol\% samples the lines are fits to the critical slowing down relation [\eq{Eq: tauc}] with the parameters given in Table \ref{Tab: dynscal}. The assumptions $E=0$ and $E=500$ yield exactly the same line. For the 0.06 vol\% sample $\Tf$ is the superparamagnetic blocking temperature defined as the maximum of $\chi''$.  
\label{Fig: tau}}
\end{figure}

Ac susceptibility data for a large set of frequencies are shown in Fig.~\ref{Fig: ac17} for the 17 vol\% sample and in Fig.~\ref{Fig: ac5} for the 5 vol\% sample.
These curves were used to extract the $T_{\rm f}(1/\omega)$ plotted in Fig. \ref{Fig: tau}.
The corresponding blocking temperatures for the 0.06 vol\% sample are plotted in the same figure as a reference for the behavior of a noninteracting system of the same particle ensemble.
A dynamic scaling analysis according to critical slowing down [Eq.~(\ref{Eq: tauc})] with the microscopic timescale given by an Arrhenius law [\eq{Eq: arrhenius}], $\tau_c(\Tf) = \tau_0 \exp(\D/\kB \Tf)(\Tf/\Tg-1)^{z \nu}$, was performed for the two concentrated samples. Two assumptions concerning the anisotropy energy was used: (i) $\D=0$, which correspond to a temperature-independent microscopic flip time; and (ii) $\D/\kB=500$~K, which is approximately the anisotropy barrier energy for a particle of average size.
The values obtained for $z \nu$, $\Tg$, and $\tau_0$ in each case are given in Table \ref{Tab: dynscal}.
The quality of the fits of the experimental data to Eq.~(\ref{Eq: tauc}) are equally good for assumptions (i) and (ii). 
In fact, in Fig.~\ref{Fig: tau}, the line corresponds to both assumptions; in the experimental temperature--frequency interval the two assumptions cannot be distinguished.
In addition, the values of $\Tg$ and the critical exponents depend quite strongly on the criterion used when determining $\Tf$.
The error bars on the exponents obtained are therefore large.
\begin{table}
\center
\caption{Parameters obtained from dynamic scaling analysis
\label{Tab: dynscal}
}
\medskip
\begin{tabular}{c c c c c}
\hline
\noalign{\smallskip}
Sample (vol\%) & $E/\kB$ (K) & $z \nu$  & $\Tg$ (K)  & $\tau_0$ (s) \\
\noalign{\smallskip}
\hline
\noalign{\smallskip}
17  & 0 & 11.4 & 48.8 & $2\cdot 10^{-8}$ \\
\noalign{\smallskip}
17  & 500 & 8.8 & 49.9 & $5\cdot 10^{-11}$ \\
\noalign{\smallskip}
5  & 0 & 10.3 & 36.0 & $2\cdot 10^{-5}$ \\
\noalign{\smallskip}
5  & 500 & 6.4 & 37.9 & $1\cdot 10^{-8}$ \\
\noalign{\smallskip}
\hline
\end{tabular}
\end{table}

\begin{figure}[htb]
\center
\includegraphics[width=0.8\textwidth]{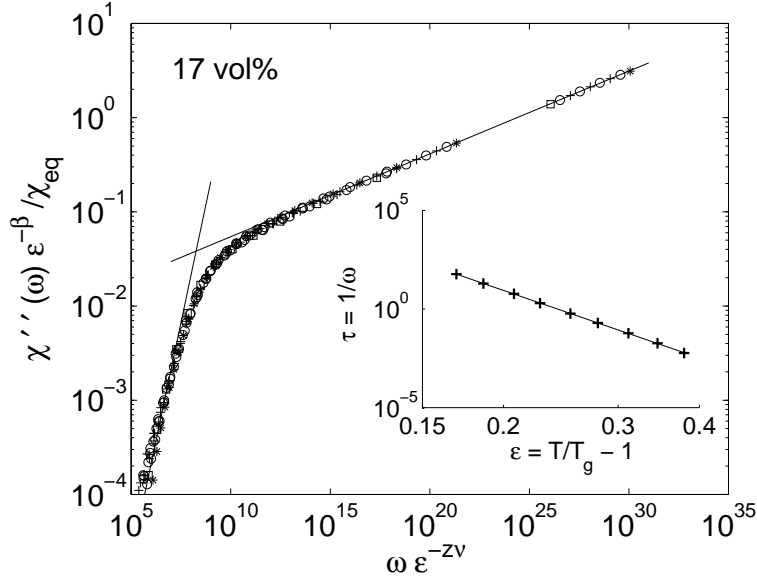}
\caption{Scaling of $\epsilon^{-\beta}\chi''(T,\omega)/\chieq$ data for $T>\Tg$ for the 17 vol\% sample. The assumption $\D=0$ is used and $\beta=1.0$. The other parameters are those of Table \ref{Tab: dynscal}. The two lines are the asymptotic behavior $G(x)\propto x$ for small values of $x$ and $G(x)\propto x^{\beta/z\nu}$ for large $x$. Inset: critical slowing down analysis on a log-log scale.
\label{Fig: dynscale}}
\end{figure}

In a full dynamics scaling analysis, the imaginary component of the  dynamic susceptibility of a spin glass is scaled according to \cite{rigaux95}
\begin{equation}
{{\chi''(T,\omega)}\over{\chi_{\rm eq}(T)}}=\epsilon^\beta
G(\omega\tau_c), \qquad T> \Tg
\label{Eq: dynscal}
\end{equation}
where $\omega=1/t$ and $G(x)$ is a scaling function. 
The asymptotic behavior of $G(x) \propto x^y$ with $y=1$ and $\beta/z\nu$ for small and large values of $x$, respectively.
It is shown in Fig. \ref{Fig: dynscale} that the $\chi''$ data for the 17 vol\% sample could be collapsed into a master curve according to this relation.
In this figure the assumption $\D=0$ is used, but an equally good collapse could be obtained by assuming $\D=500$~K (and changing $\epsilon^\beta$ to $\tau_m^{\beta/z\nu}$). 
The critical exponent $\beta =1.0 \pm 0.3$.
For the 5 vol\% sample, the dynamic susceptibility could not be scaled according to \eq{Eq: dynscal}.
In addition, if data for larger values of $\tau$ (obtained from ZFC relaxation data) are included in the critical slowing down analysis, deviations from the power law is observed \cite{hanetal2002}.
To conclude, the dynamic scaling analysis indicates that the 5 vol\% sample does not exhibit a thermodynamic phase transition although it clearly exhibits spin glass dynamics.
Because of the temperature dependence of $\tau_m$, a static scaling analysis as performed in Ref. \cite{jonsvehan98}  is a crucial additional tool to disclose a possible spin glass phase transition in interacting nanoparticle systems.

\begin{figure}[htb]
\center
\includegraphics[width=0.8\textwidth]{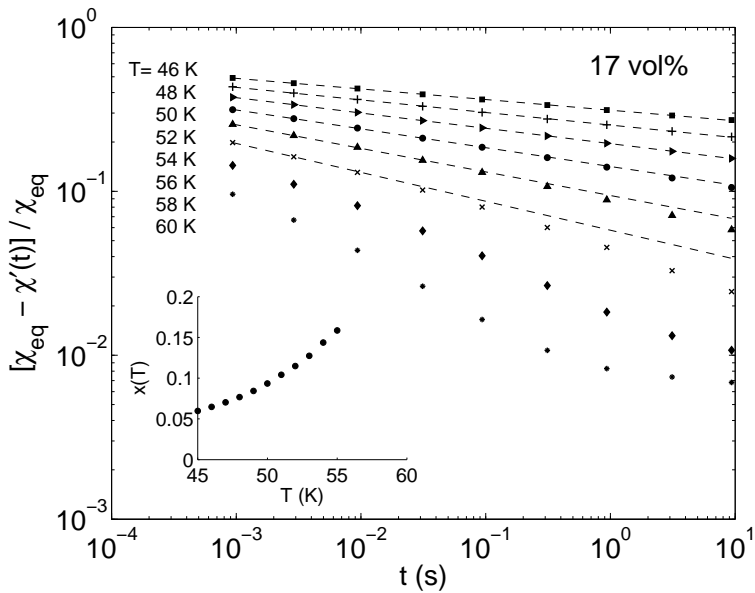}
\caption{$[\chieq-\chi'(t)]/\chieq$ vs time for the 17 vol\% sample.
Inset: temperature dependence of the exponent $x$ in Eq.~(\ref{Eq: q}).
\label{Fig: q17}}
\end{figure}

The time dependence of the dynamic correlation function $q(t)$ was investigated numerically on the Ising EA model by Ogielski \cite{ogielski85}.
An empirical formula for the decay of $q(t)$ was proposed as a combination of a power law at short times and a stretched exponential at long times
\be
q(t) = c t^{-x} e^{-w t^y} \,,
\label{Eq: q}
\ee
where $c$, $x$, $w$, and $y$ are temperature-dependent parameters.
$q(t)$ follows a pure power-law behavior below $\Tg$.
It was shown in Ref. \cite{gunetal88} that $[\chieq -\chi'(t)]/\chieq$ measured on the \ising spin glass sample also behaves according to Eq.~\ref{Eq: q}.
In Fig.~\ref{Fig: q17} $[\chieq -\chi'(t)]/\chieq$ is shown for the 17 vol\% sample at temperatures around $\Tg$. 
The temperature dependence of the exponent $x$ is shown in the inset.
The behavior of the 17 vol\% sample shown here is similar to that observed in the numerical simulation on the EA Ising model \cite{ogielski85} and to  the experimental result obtained for the \ising spin glass \cite{gunetal88}.
However, the stretched--exponential behavior is less pronounced than for the \ising spin glass. One reason for this is that the investigated timescales $t/\tau_m$ are shorter for the nanoparticle sample because of the larger value of $\tau_m$. 
In the numerical simulation, performed on even shorter timescales, it was possible to observe the stretched--exponential behavior by investigating $q(t)$ far from $\Tg$. 
The resolution of our experimental data does not allow such an investigation.

\subsubsection{Dynamics in a Field}

\begin{figure}[bht]
\center
\includegraphics[width=0.78\textwidth]{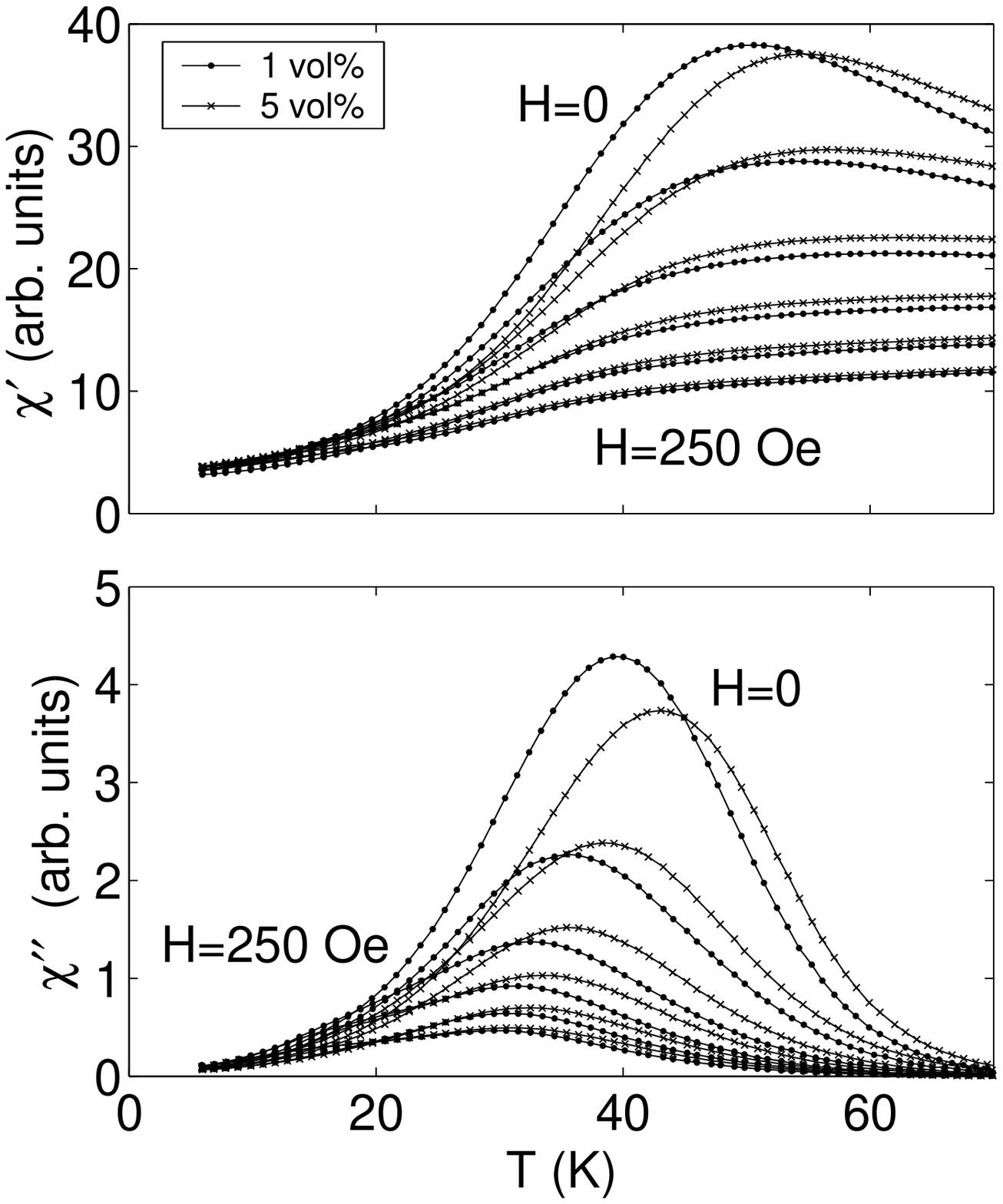}
\caption{ac susceptibility versus temperature for different
superimposed dc fields; $H$ = 0, 50, 100, 150, 200, 250 Oe. $\omega/2\pi=125$~Hz.
\label{Fig: acdc}}
\end{figure}

All measurements presented so far are performed with a probing field in the linear response regime and without an external bias field.
We will now investigate the effect of a nonzero bias field on the nonequilibrium dynamics (still probing the system with a weak magnetic field in the linear response regime).
The question as to whether a  spin-glass-like phase exists under a finite field in a strongly  interacting nanoparticle system will not be addressed.
Figure \ref{Fig: acdc} shows the ac susceptibility as a function of temperature with bias dc fields in the range 0 -- 250 G for  1 and a 5 vol\% samples.
The magnetic field affects the superparamagnetic behavior of individual spins (as discussed in section \ref{Chap: nano}) as well as the nonequilibrium dynamics.
It can be seen in the figure that the dynamic response of the two systems in low fields is rather different, while with increasing bias field the dynamic response of the two systems becomes more and more similar.
This indicates that the effects of dipolar interaction are suppressed by a sufficiently strong magnetic field.

\begin{figure}[bht]
\center
\includegraphics[width=0.78\textwidth]{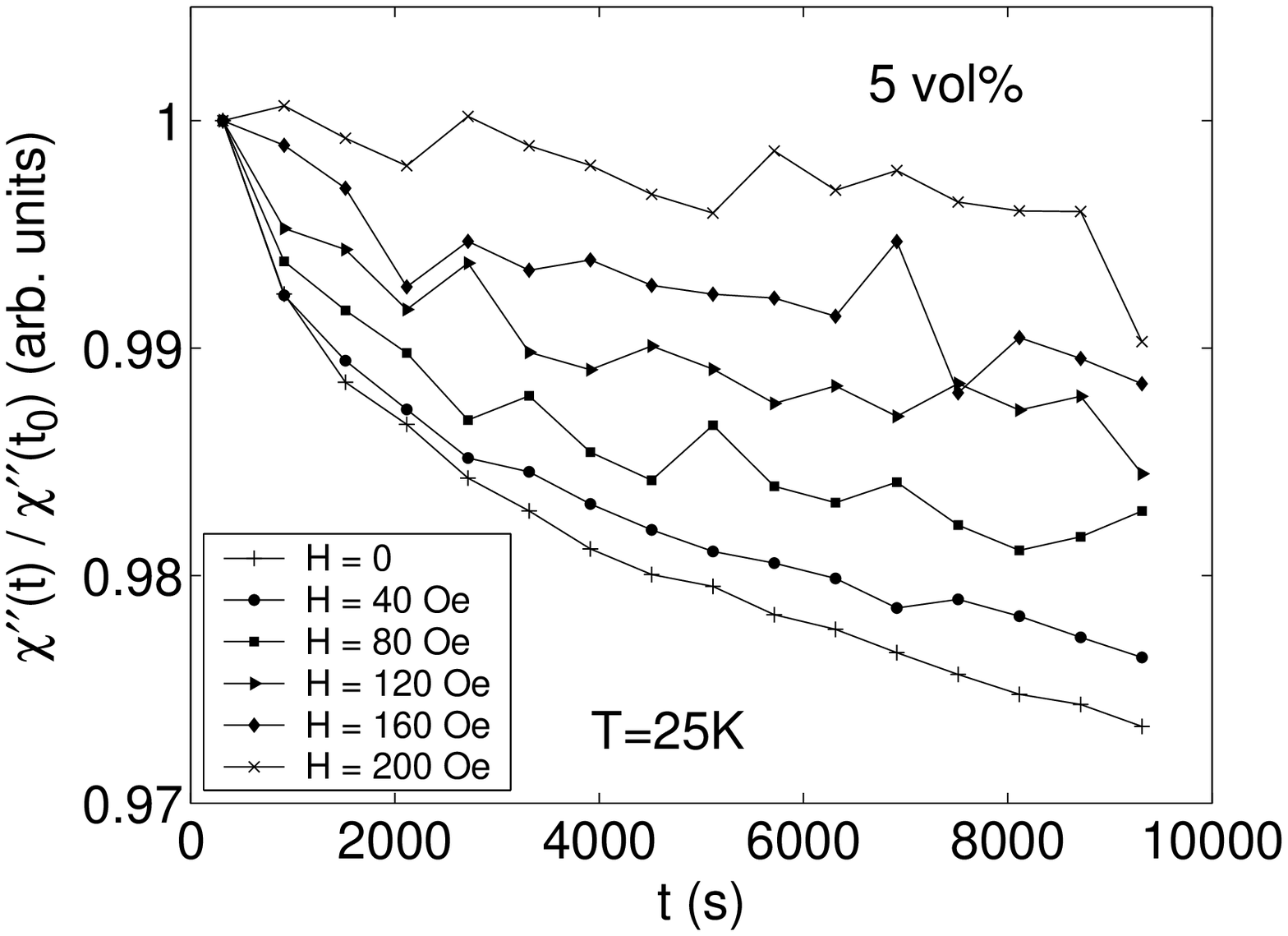}
\caption{ac susceptibility versus time for different
superimposed dc fields; $H$ = 0, 40, 80, 120, 160, 200 Oe. $\omega/2\pi=510$~Hz.
\label{Fig: acrelaxH}}
\end{figure}

The effect of an external field on the glassy dynamics can be studied by recording the ac susceptibility as a function of time for different bias fields.
$\chi''$ normalized by its value just after the quench, $\chi''(t_0)$, is shown in Fig.~\ref{Fig: acrelaxH} for the 5 vol\% sample.
It can be seen in this figure that the ac relaxation diminishes with increasing bias field. For the highest field, almost no relaxation exists.
The field has in addition the effect that it makes the measurement more noisy.
The effects of the field is also temperature-dependent and the ac susceptibility is more affected by a field at high temperatures.
A qualitatively similar result was found for the (Fe$_{0.15}$Ni$_0.85$)$_{75}$P$_{16}$B$_6$Al$_3$ spin glass \cite{jonetal2001PRB}.

\subsection{Discussion}

We have seen that the magnetic properties of a strongly interacting nanoparticle system are of spin-glass-like nature, and hence very different from the superparamagnetic behavior of noninteracting systems and weakly interacting spins discussed in section \ref{Chap: nano}.
Any model for interparticle interaction based on a modified superparamagnetic behavior \cite{dorbesfio88,mortro94,jongar2001EPL,luietal2002} will therefore fail to describe the dynamics of a strongly interacting nanoparticle system. 

The value of the dipolar coupling parameter $h_d$ [\eq{Eq: hd}] determines the strength of the dipolar interaction, but the width of the distribution of energy barriers is equally important for the dynamic properties. For example, Jonsson et al. studied \cite{jonnorsve98} a $\gamma$-Fe$_2$O$_3$ nanoparticle sample  that had a value of $\xid$ comparable to the 17 vol\% sample investigated here, but with a much broader energy barrier distribution.
That sample showed glassy dynamics, but it did not exhibit a spin glass phase transition.

It would be interesting to examine the nonequilibrium dynamics of a strongly interacting system for different concentrations (different dipolar coupling strengths). Such a study could give more detailed insight into how the aging and rejuvenation phenomena depend on the interaction strength and the experimental timescale.

\section{Summary and Conclusion}
\label{Chap: conclusion}

The effects of interparticle dipolar interaction in magnetic nanoparticle systems have been investigated.
For weak dipolar coupling strengths equilibrium quantities can be calculated using thermodynamic perturbation theory, treating the anisotropy exactly and the dipolar interaction perturbatively.
Such an approach is always valid at sufficiently high temperatures, but it is valid around the blocking temperature only in the case of weak dipolar coupling.
In a simple model, the relaxation rate is modified by the dipolar interaction in the same way as by a field, where the field components of the dipolar field can be  calculated using thermodynamic perturbation theory \cite{jongar2001EPL}.
The dipolar field also plays an important role for quantum tunneling of the magnetization of crystals of magnetic molecules  \cite{wernsdorfer2001}.

Superparamagnetic blocking depends strongly on the damping parameter in the case of weak--medium damping due to the transverse component of the dipolar field \cite{garetal99}. 
Any energy-barrier-based model overlooks that damping dependence and can therefore only be valid in the overdamped case.
The effect of a magnetic field is to decrease the relaxation time. 
Hence, in the case of weak interparticle interaction, a decrease of the blocking temperature is predicted and also observed in high-frequency measurements, such as M{\"o}ssbauer spectroscopy \cite{mortro94} and Langevin dynamics simulations \cite{bergor2001}.
However, for  stronger interparticle interaction, the dipolar interaction does not modify but creates energy barriers, and hence the  blocking temperature increases with increasing interaction strength.
A blocking temperature that increases with the interaction strength is commonly observed in magnetization measurements.

The existence of glassy dynamics in dense frozen ferrofluids, specifically strongly interacting nanoparticle systems with randomness in the particle positions and anisotropy axes, has been evidenced by experimental techniques developed in the study of conventional spin glasses.
Hence, the dynamics of such systems is radically different from simple superparamagnetic blocking.
The nonequilibrium dynamics observed in strongly interacting nanoparticle systems exhibits qualitatively similar aging, memory, and rejuvenation effects as ordinary spin glasses, but the aging and rejuvenation effects are weaker.
The differences observed can be explained at least qualitatively  by the longer microscopic relaxation time of a magnetic moment compared to an atomic spin.
Only strongly interacting nanoparticle systems with a narrow anisotropy barrier distribution have been shown to exhibit a phase transition to a low-temperature spin glass phase.

\section*{Acknowledgments}

The author is indebted to J. L. Garc{\'{\i}}a-Palacios, M. F. Hansen  and  P. Nordblad for their contribution to the work presented in this article. 
The author also acknowledges  
H. Aruga Katori, S. Felton, A. Ito, T. Jonsson,  R. Mathieu, P. Svedlindh, and H. Yoshino
for collaboration and
I. A. Campbell, H. Mamiya, T. Sato, H. Takayama, and E. Vincent
for helpful discussions.

This work has partially been financed by the Swedish Research Council (VR).

\begin{appendix}

\newcommand{\Ra}{\frac{S_{4}}{35}-\frac{2S_{2}}{21}+\frac{1}{15}}
\newcommand{\Rb}{\frac{1}{7}(S_{2} - S_{4})}
\newcommand{\Ei}{{\bf \Gamma}_{i}}
\newcommand{\Ej}{{\bf \Gamma}_{j}}
\newcommand{\Ek}{{\bf \Gamma}_{k}}
\newcommand{\Di}{{\bf \Lambda}_{i}}
\newcommand{\nj}{\vec{n}_j}
\newcommand{\Ga}{{\bf \Gamma}}
\newcommand{\Gb}{{\bf \Lambda}}

\section{Thermodynamic Perturbation Theory}
\label{App: TPT}

\subsection{Expansion of the Boltzmann Distribution in the Dipolar Coupling Parameter}
\label{App: expTPT}
Thermodynamic perturbation theory is used to expand the Boltzmann distribution in the dipolar interaction, keeping it exact in the magnetic anisotropy (see Sec. \ref{Sec: TPT}). 
A convenient way of performing the expansion in
powers of $\xid$ is to introduce the Mayer functions $f_{ij}$
defined by $1+\fij=\exp(\xid\wij)$, which permits us to write the
exponential in the Boltzmann factor as
\begin{equation}
\exp(-\beta \Hamil)
=
\exp(-\beta \Ea)\prod_{i>j}(1+\fij).
\label{expBoltz}
\end{equation}
Expanding the product to second order in the $\fij$ gives
\begin{equation}
\label{prod:fij}
\prod_{i>j}(1+\fij)
=
1
+
\xid
\G_{1}
+
\case{1}{2}
\xid^2
\G_{2}
+
O(\xid^3)
,
\end{equation}
where \cite{ros-lax52}
\begin{eqnarray}
\label{g1}
\G_{1}
&=&
\sum_{i>j}\wij,
\\
\label{g2}
\G_{2}
&=&
\sum_{i>j} \wij^2 + \sum_{i>j} \sum_{k>l} \wij \wkl \qikjl \qiljk,
\end{eqnarray}
and the symbol $\qikjl$ annihilates terms containing duplicate pairs:
$\qikjl =
\case{1}{2}
(2-\delta_{ik}-\delta_{jl})
(1+\delta_{ik})(1+\delta_{jl})$.

To obtain the average of any quantity $\obs$, we introduce the
expansion (\ref{prod:fij}) in both the numerator and denominator of
$\la\obs\ra
=
\int\!\dW\obs\,\exp (-\beta \Hamil)
/\int\!\dW\exp (-\beta \Hamil)$,
and work out the expansion of the quotient, getting 
\be
\la\obs\ra
\simeq
\la\obs\ra_{\rm a}
+
\xid
\la\obs\,\G_{1} \ra_{\rm a}
+\case{1}{2}
\xid^2
\big[
\la\obs\,\G_{2} \ra_{\rm a}
-
\la\obs\ra_{\rm a} \la \G_{2} \ra_{\rm a}
\big].
\label{avg:g1=0}
\ee
Here we have utilized the fact that the  single-spin anisotropy has inversion symmetry
[$\Wa(-\ei)=\Wa(\ei)$] in the absence of a  bias
field and that $\la \G_{1} \raa = \la \sum_{i>j} \wij \raa = 0$ since a dipole does not interact with itself.

We have now obtained expressions for $F_1$ and $F_2$ in Eq.~(\ref{W:approx}):
\be
F_1 = G_1 \, , \qquad F_2 = G_2 -\la G_2 \raa \, .
\ee
To complete the calculation, we need to obtain averages of low-grade
powers of $\e$ weighted by the noninteracting distribution (moments),
which is the only place where one needs to specify the form of $\Ea$.
In the next section we will do that for systems with axially symmetric anisotropy.

\subsection{Averages Weighted with an  Axially Symmetric Boltzmann Factor}
\label{App:alg}

Assuming an axially symmetric potential, the anisotropy energy of $\Ea(\e \cdot \n)$ will be an {\em even} function of the longitudinal component of the magnetic moment $\e \cdot \n$.
The averages we need to calculate are all products of the form $I_m
=
\left\langle
\prod_{n=1}^{m}(\vec{c}_n \cdot \e)
\right\rangle_{\rm a}$,
where the $\vec{c}_{n}$ are arbitrary constant vectors.
Introducing the polar and azimuthal angles of the spin
$(\vartheta,\varphi)$, we can write $I_{m}$ as
\[
I_m
=
\frac{\int_0^{2\pi} d\varphi \int_0^{\pi} d\vartheta \sin\vartheta
\prod_{n=1}^{m}(\vec{c}_n \cdot \e) \exp[-\beta \Ea(\e\cdot\n)]}
{\int_0^{2\pi} d\varphi \int_0^{\pi} d\vartheta \sin\vartheta
\exp[-\beta \Ea(\e\cdot\n)]}
\;.
\]
For odd $m$, $I_m$ is an integral of an odd function over a
symmetric interval and hence $I_m=0$.
To calculate the susceptibility and specific heat to second order in
$\xid$, we require $I_{2}$ and $I_4$, which will be
calculated using symmetry arguments similar to those employed to
derive the $\sigma=0$ unweighted averages (see, e.g., 
Ref.~\cite{mathews-walker}).

Note that $I_{2}$ is a scalar bilinear in $\vec{c}_{1}$ and
$\vec{c}_{2}$.
The most general scalar with this property that can be constructed
with the vectors of the problem ($\vec{c}_{1}$, $\vec{c}_{2}$, and
$\n$) has the form
\[
I_{2}
=
A \, \vec{c}_{1} \cdot \vec{c}_{2}
+
B \,(\vec{c}_{1} \cdot \n)(\vec{c}_{2} \cdot \n)
\;.
\]
To find the coefficients $A$ and $B$, one chooses particular values for
the $\vec{c}_n$: 
\begin{enumerate}
\item[(i)] 
If $\vec{c}_{1} \parallel
\vec{c}_{2} \perp \n$ then $I_{2} = A$.
Thus, setting $\n=\hat{z}$ and $\vec{c}_{1}=\vec{c}_{2}=\hat{x}$, one
has $\e\cdot\n=\cos\vartheta\equiv z$ and $( \vec{c}_{1} \cdot
\e)(\vec{c}_{2} \cdot \e)=(1-z^{2})\cos^2\varphi$, so the integral reads
\begin{eqnarray*}
A
&=&
\frac{
\int_0^{2\pi} d\varphi \cos^2\varphi
\int_{-1}^{1} dz
(1-z^{2})
\exp[-\beta \Ea(z)]
}{
\int_0^{2\pi} d\varphi
\int_{-1}^{1} dz
\exp[-\beta \Ea(z)]
}
\\
&=&
\case{1}{2}
\big[
1-\big\langle z^{2}\big\rangle_{\rm a}
\big]
=
\Pa
\;,
\end{eqnarray*}
where $S_{2}=\la P_{2}(z) \ra_{\rm a}$ is the average of the
second Legendre polynomial
$P_{2}(z)=\frac{1}{2}\left(3\,z^{2}-1\right)$ over the noninteracting
distribution.
\item[(ii)] 
If $\vec{c}_{1}\parallel\vec{c}_{2}\parallel\n$, then
$I_{2}=A+B$.
Putting $\n=\vec{c}_{1} = \vec{c}_{2} = \hat{z}$ the integral is given
by
\[
A+B
=
\frac{
\int_{-1}^{1} dz\,
z^{2}
\exp[-\beta \Ea(z)]
}
{
\int_{-1}^{1} dz
\exp[-\beta \Ea(z)]
}
=
\big\langle
z^{2}
\big\rangle_{\rm a}
=
\frac{1+2S_{2}}{3}
\;.
\]
\end{enumerate}
Therefore, since
$I_{2}=\la ( \vec{c}_{1} \cdot \e)(\vec{c}_{2} \cdot \e)\ra_{\rm a}$,
we get the following for the second-order moment:
\begin{equation}
\la ( \vec{c}_{1} \cdot \e)(\vec{c}_{2} \cdot \e) \ra_{\rm a}
=
\Pa \, \vec{c}_{1} \cdot \vec{c}_{2}
+ S_{2} (\vec{c}_{1} \cdot \vec{n})(\vec{c}_{2} \cdot \vec{n})
\;.
\label{alg2}
\end{equation}
We can similarly calculate $I_4$ by constructing the
most general scalar fulfilling certain properties, getting
{\small
\begin{eqnarray}
\langle
(\vec{c}_{1} \cdot \e)
(\vec{c}_{2} \cdot \e)
(\vec{c}_{3} \cdot \e)
(\vec{c}_{4} \cdot \e)
\rangle_{\rm a}
&=&
\Delta_{4} \,
[ (\vec{c}_{1} \cdot \vec{c}_{2})(\vec{c}_{3} \cdot \vec{c}_{4})+
(\vec{c}_{1} \cdot \vec{c}_{3})(\vec{c}_{2} \cdot \vec{c}_{4})\nonumber\\
&& +
(\vec{c}_{1} \cdot \vec{c}_{4})(\vec{c}_{2} \cdot \vec{c}_{3}) ]
\nonumber\\
& & +
\Delta_{2} \,
[(\vec{c}_{1} \cdot \vec{c}_{2})(\vec{c}_{3} \cdot \vec{n})
(\vec{c}_{4} \cdot \vec{n})
+
(\vec{c}_{1} \cdot \vec{c}_{3})(\vec{c}_{2} \cdot \vec{n})
(\vec{c}_{4} \cdot \vec{n})
\nonumber\\
& & 
+(\vec{c}_{1} \cdot \vec{c}_{4})(\vec{c}_{2} \cdot \vec{n})
(\vec{c}_{3} \cdot \vec{n})
+
(\vec{c}_{2} \cdot \vec{c}_{3})(\vec{c}_{1} \cdot \vec{n})
(\vec{c}_{4} \cdot \vec{n})
\nonumber\\
& & 
+(\vec{c}_{2} \cdot \vec{c}_{4})(\vec{c}_{1} \cdot \vec{n})
(\vec{c}_{3} \cdot \vec{n})
+
(\vec{c}_{3} \cdot \vec{c}_{4})(\vec{c}_{1} \cdot \vec{n})
(\vec{c}_{2} \cdot \vec{n})]
\nonumber\\
& &
+ S_{4} (\vec{c}_{1} \cdot \vec{n})(\vec{c}_{2} \cdot \vec{n})
(\vec{c}_{3} \cdot \vec{n})(\vec{c}_{4} \cdot \vec{n}),
\label{alg4}
\end{eqnarray}
}
where $\Delta_{2}$ and $\Delta_{4}$ are combinations of the first
$S_l(\sigma)$
\begin{equation}
\label{deltas}
\Delta_{2}
=
\Rb
,
\qquad
\Delta_{4}
=
\Ra
\;.
\end{equation}
Therefore, Eq.\ (\ref{alg4}) involves $S_{2}$ as well as
$S_{4}=\la P_{4}(z)\ra_{\rm a}$, the average of the fourth
Legendre polynomial
$P_{4}(z)=\frac{1}{8}\left(35\,z^{4}-30\,z^{2}+3\right)$ with respect
to $\Wa$.

Finally, introducing the following tensor and scalar shorthands
\begin{eqnarray}
\label{tensorGa}
\Ga
&=&
\Pa \I + S_{2} \, \n\,\n
\;,
\\
\label{tensorGb}
\Gb
&=&
\sqrt{\Delta_{4}}\, \I + \frac{\Delta_{2}}{\sqrt{\Delta_{4}}} \,\n\,\n
\;,
\quad
\Omega
=
S_{4} - 3\frac{\Delta_{2}^2}{\Delta_{4}}
\;,
\end{eqnarray}
where $\I$ is the identity tensor, the results for the moments can
compactly be written as
\begin{eqnarray}
\la ( \vec{c}_{1} \cdot \e)(\vec{c}_{2} \cdot \e) \ra_{\rm a}
&=&
(\vec{c}_{1} \cdot \Ga \cdot \vec{c}_{2})
\\
\langle
(\vec{c}_{1} \cdot \e)
(\vec{c}_{2} \cdot \e)
(\vec{c}_{3} \cdot \e)
(\vec{c}_{4} \cdot \e)
\rangle_{\rm a}
&=&
(\vec{c}_{1} \cdot \Gb \cdot \vec{c}_{2})
(\vec{c}_{3} \cdot \Gb \cdot \vec{c}_{4})
\nonumber\\
& &
{}+(\vec{c}_{1} \cdot \Gb \cdot \vec{c}_{3})(\vec{c}_{2} \cdot \Gb \cdot
\vec{c}_{4})
\nonumber\\
& &
{}+(\vec{c}_{1} \cdot \Gb \cdot \vec{c}_{4})(\vec{c}_{2} \cdot \Gb \cdot
\vec{c}_{3})
\\
& &
{}+\Omega (\vec{c}_{1} \cdot \n)(\vec{c}_{2} \cdot \n)
(\vec{c}_{3} \cdot \n)(\vec{c}_{4} \cdot \n)
\;,\nonumber
\end{eqnarray}
which facilitates the manipulation of the
observables.

The quantities $S_{l}$ are calculated in the case of uniaxial anisotropy in Appendix \ref{App: Sl}.
Note finally that in the isotropic limit ($S_l\to0$), Eqs.~(\ref{alg2})
and (\ref{alg4}) reduce to the known moments for the isotropic
distribution \cite{ros-lax52,mathews-walker}
\begin{eqnarray}
\label{alg2:iso}
\langle ( \vec{c}_{1} \cdot \e)(\vec{c}_{2} \cdot \e)
\rangle_{{\rm iso}}
&=&
\case{1}{3} \, \vec{c}_{1} \cdot \vec{c}_{2}
\;,
\\
\label{alg4:iso}
\langle
(\vec{c}_{1} \cdot \e)
(\vec{c}_{2} \cdot \e)
(\vec{c}_{3} \cdot \e)
(\vec{c}_{4} \cdot \e)
\rangle_{{\rm iso}}
&=&
\case{1}{15}
[
(\vec{c}_{1} \cdot \vec{c}_{2})(\vec{c}_{3} \cdot \vec{c}_{4})
\nonumber\\
& &
+
(\vec{c}_{1} \cdot \vec{c}_{3})(\vec{c}_{2} \cdot \vec{c}_{4})
\nonumber\\
& &
+
(\vec{c}_{1} \cdot \vec{c}_{4})(\vec{c}_{2} \cdot \vec{c}_{3})]
\;.
\end{eqnarray}
These expressions are formally identical to those for the
average of a quantity involving the anisotropy axes $\nii$, when
these are distributed at random
$\frac{1}{N}
\sum_{i}
f(\nii)
\rightarrow
\int
\frac{d^{2}\n}{4\pi}
f(\n)
\equiv
\overline{f}$.
For instance, for arbitrary $\n$-independent vectors $\vec{v}_{1}$ and
$\vec{v}_{2}$, we have
\begin{equation}
\label{alg2:n}
\frac{1}{N}
\sum_{i}
(\vec{v}_{1}\cdot\nii)(\vec{v}_{2}\cdot\nii)
\longrightarrow
\overline{(\vec{v}_{1}\cdot\n)(\vec{v}_{2}\cdot\n)}
=
\case{1}{3} \, \vec{v}_{1}\cdot\vec{v}_{2}
\;.
\end{equation}

\subsection{General Formulae for the Coefficients of Susceptibility}
\label{App:linsusc}

The general expression for the equilibrium linear susceptibility is
given by Eq.~(\ref{xeq-TPT}) with the following expressions for the
coefficients
\begin{eqnarray}
\coeff_{0}
&=&
\frac{1}{N}\sum_{i} \h\cdot\Ei\cdot\h
\\
\coeff_{1}
&=&
\frac{1}{N}\sum_{i} \sum_{j \ne i}
\h\cdot\left(\Ei \cdot \Gij \cdot \Ej\right)\cdot\h
\\ \nonumber
\coeff_{2}
&=&
- \frac{2}{N} \sum_{i} \sum_{j \ne i}
\h\cdot\left(\Ei \cdot \Gij \cdot \Ej \cdot \Gij \cdot \Ei\right)\cdot\h
\\ \nonumber
& &
{}+
\frac{2}{N} \sum_{i} \sum_{j \ne i}\sum_{k \ne j}
\h\cdot\left( \Ei \cdot \Gij \cdot \Ej \cdot \Gjk \cdot \Ek \right)
\cdot\h
\\ \nonumber
& &
{}+ \frac{1}{N}\sum_{i} \sum_{j \ne i}
\bigg\{
\frac{1-S_{2}}{r_{ij}^{6}}
\Big[ (\h\cdot\Di\cdot\h)(\vij \cdot \Di \cdot \vij)
\\ \nonumber
& &
\qquad\qquad\qquad
\qquad\qquad
{}+2(\h\cdot\Di \cdot \vij)^2
\\ \nonumber
& &
\qquad\qquad\qquad
\qquad\qquad
{}+
\Omega (\h\cdot\nii)^2(\nii \cdot \vij)^2\Big]
\\ \nonumber
& &
\qquad\qquad\qquad
{}+S_{2}
\Big[
(\h\cdot\Di\cdot\h)
(\nj \cdot \Gij \cdot \Di \cdot \Gij \cdot \nj)
\\ \nonumber
& &
\qquad\qquad\qquad
\qquad\quad
{}+
2 (\h\cdot\Di \cdot \Gij \cdot \nj)^2
\\ \nonumber
& &
\qquad\qquad\qquad
\qquad\quad
{}+
\Omega (\h\cdot\nii)^2(\nii \cdot \Gij \cdot \nj)^2
\Big]
\bigg\}
\\ \nonumber
& &
{} - \frac{1}{N}\sum_{i} \sum_{j \ne i} (\h\cdot\Ei\cdot\h)
\bigg[
\frac{1-S_{2}}{r_{ij}^{6}}
(\vij \cdot \Ei \cdot \vij)
\\ 
& &
\qquad\qquad\qquad
\qquad\qquad\quad
{}+S_{2} (\nj \cdot \Gij \cdot \Ei \cdot \Gij \cdot \nj)
\bigg]
\end{eqnarray}
where $\Gij$, $\rij$ and $\vij$ are defined in Eq.\ (\ref{Gij}), and
$\Ga$, $\Gb$, and $\Omega$ in Eqs.\ (\ref{tensorGa}) and (\ref{tensorGb})
and also involve the $S_l(\sigma)$.

When calculating these coefficients, the same type of averages appear as
in the isotropic case (see Refs.~\cite{vanvle37,ros-lax52} for
details of the calculation) and with the same multiplicities.
The only difference is the weight function and hence the formulas
required to calculate those averages [Eqs.~(\ref{alg2}) and
(\ref{alg4}) instead of Eqs.~(\ref{alg2:iso}) and (\ref{alg4:iso})].

\subsection{General Formula for the Coefficient $b_2$ of Specific Heat}
\label{App:cV}
 
In the general expression (\ref{cVexp}) for the specific heat the
coefficient $b_{0}$ is given by Eq.\ (\ref{b0gen}), while $b_{2}$ reads
\begin{eqnarray}
Nb_2 
&=&
\case{1}{3} \big\{ 2(1-S_2) -4\sigma S_2' -\sigma^2S_2'' \big\}
 \sum_{i}\sum_{j \ne i} r_{ij}^{-6} 
\nonumber \\
& &
+\case{1}{2}
\big\{
2S_2(1-S_2) + 4\sigma S_2'(1-2S_2)
\nonumber\\
& &
\qquad
{}+ \sigma^2 [S_2''(1-2S_2) - 2 (S_2')^2]
\big\} 
\nonumber\\
& &
\quad
\times
\sum_{i} \sum_{j \ne i} r_{ij}^{-6}
\big[(\vij \cdot \nii)^2+(\vij \cdot \nj)^2\big]
\nonumber \\
& &
+
\big\{ S_2^2 + 4\sigma S_2 S_2' + \sigma^2 [S_2S_2'' + (S_2')^2] \big\}
\sum_{i}\sum_{j \ne i} (\nii \cdot \Gij \cdot \nj)^2 \; ,
\label{b2gen}
\end{eqnarray}
where $f'=df/d\sigma$.
General formulae for $S_l'$ in the case of uniaxial anisotropy are given in Appendix \ref{App: Sl}.


\subsection{Dipolar Field}
\label{App:df}

The dipolar field averages we want to calculate are $\langle \xi_{i,\|}^{2}\rangle 
= \langle (\vec{\xi}_{i} \cdot \nii)^{2}\rangle$ and  $\langle \xi_{i,\perp}^{2}\rangle 
= \langle {\xi}_{i}^{2}\rangle - \langle \xi_{i,\|}^{2}\rangle $.
The general expressions for these quantities are \cite{jongar2001EPL}
\begin{eqnarray}
\label{h:para}
\big\langle
\xi_{i,\|}^{2}
\big\rangle
&=&
\frac{\xid^{2}}{3}
\sum_{j}
\big[
\left(1-S_{2}\right)
\,
(\nii\cdot\Gij\cdot\Gij\cdot\nii)
\nonumber\\
&& 
+3S_{2} \, (\nii\cdot\Gij\cdot\nj)^{2}\big]
\;,
\\
\label{h:perp}
\big\langle
\xi_{i,\perp}^{2}
\big\rangle
&=&
\frac{\xid^{2}}{3}
\sum_{j}
\big[
6r_{ij}^{-6}
+
3S_{2} \, r_{ij}^{-3} (\nj\cdot\Gij\cdot\nj)
\nonumber\\
& & 
-\left(1-S_{2}\right)
\,
(\nii\cdot\Gij\cdot\Gij\cdot\nii)
\nonumber \\
& & 
-3S_{2} \, (\nii\cdot\Gij\cdot\nj)^{2}\big]
\;.
\end{eqnarray}

Since we want to use the dipolar fields in order to examine the blocking behavior (see Sec. \ref{Sec: relax-int}), one may wonder about the validity of these field averages below the
superparamagnetic blocking, where the spins are not in complete
equilibrium.
However, since at those temperatures the spins are still in
quasiequilibrium confined to one of the two wells, we can repeat the
derivation of the algorithm (\ref{alg2}) restricting the phase space
for integration to one well.
In this case, averages of the form $\la\e\cdot\vec{v}_{1}\raa$ do not
vanish, and should be considered together with
$\la (\e\cdot\vec{v}_{1})(\e\cdot\vec{v}_{2}) \raa$, which, being even
in $\e$, is not modified.
The extra terms associated with $\la \e\cdot\vec{v}_{1} \raa$
will, however, vanish if the overall state is demagnetized, and 
Eqs.\ (\ref{h:para}) and (\ref{h:perp}) are recovered.

\section{$S_l$ for Uniaxial Anisotropy}
\label{App: Sl}

The thermodynamical average $S_l(\sigma)$ over the Legendre polynomials $P_l$ occur in the expressions for the susceptibilities, the specific heat, and the dipolar fields in Sec. \ref{Sec: teq}.
For uniaxial anisotropy these averages read
\be
S_l(\sigma)=\la P_l\ra_a 
= \frac{1}{\Z_a} \int_{-1}^{1} dz P_l(z) e^{\sigma z^2} \, .
\ee
In particular, $S_0 = 1$ and $S_2=\frac{1}{2}\la 3 z^2 -1 \ra_a$ can be written
\be
S_2 = \frac{3}{2}\left(\frac{e^\sigma}{\sigma \Z_a}-\frac{1}{2\sigma}\right) -\frac{1}{2} \, .
\ee
The one-spin partition function $\Z_a=\int_{-1}^{1}d z\,\exp(\sigma
z^{2})$ can be written in terms of {\em error\/} functions of real and
``imaginary'' argument as
\begin{equation}
\label{Z}
\Z_a
=
\left\{
\begin{array}{ll}
\sqrt{\pi/\sigma}\,{\rm erfi}(\sqrt{\sigma})
,
&
\sigma>0
\\
\sqrt{\pi/|\sigma|}\,{\rm erf}(\sqrt{|\sigma|})
,
&
\sigma<0
\end{array}
\right.
\end{equation}
The less familiar ${\rm erfi}(x)$ is related to the Dawson integral
$D(x)$, so in the easy-axis case one can write
$\Za=(2e^{\sigma}/\sqrt{\sigma})D(\sqrt{\sigma})$ and compute $D(x)$
with the subroutine DAWSON of Ref.\ \cite{recipes}.

For $l>2$, the  $S_{l}$ can be computed using the following homogeneous
three-term recurrence relation \cite{kalcof97}:
\begin{equation}
\label{Xlm:EOM:uniaxial:m=0:stat}
\bigg[
1
-
\frac{2\sigma}{(2l-1)(2l+3)}
\bigg]
S_{l}
-
\frac{2\sigma}{2l+1}
\bigg[
\frac{l-1}{2l-1}
S_{l-2}
-
\frac{l+2}{2l+3}
S_{l+2}
\bigg]
=
0
\;,
\end{equation}
The derivative of any $S_l$ can be computed by means of the differential recurrence relation \cite{jongar2001PRB}
\begin{eqnarray}
\label{dSldsig}
S_l'=
\frac{d S_{l}}{d\sigma}
&=&
\frac
{(l-1)l}
{(2l-1)(2l+1)}
S_{l-2}
+
\frac
{2l(l+1)}
{3(2l-1)(2l+3)}
S_{l}
\nonumber\\
& &
{}+
\frac
{(l+1)(l+2)}
{(2l+1)(2l+3)}
S_{l+2}
-
\frac{2}{3}
S_{2}
S_{l}
\;.
\end{eqnarray}

The approximate behavior of $S_{2}$ and $S_4$ for weak
($|\sigma|\ll1$) and strong ($|\sigma|\gg1$) anisotropy are
\begin{eqnarray}
\label{S2exp}
S_{2}(\sigma)
&=&
\left\{ \begin{array}{ll}
\frac{2}{15}\sigma+\frac{4}{315}\sigma^2 + \cdots
&
|\sigma| \ll 1
\\
1 - \frac{3}{2\sigma}-\frac{3}{4\sigma^2} + \cdots
&
\sigma \gg 1
\\
-\frac{1}{2}(1+\frac{3}{2\sigma}) +\cdots
&
\sigma \ll -1
\end{array}
\right.,
\\
\label{S4exp}
S_{4}(\sigma)
&=&
\left\{
\begin{array}{ll}
\frac{4}{315}\sigma^2 + \cdots
&
|\sigma| \ll 1
\\
1 - \frac{5}{\sigma}+\frac{25}{4\sigma^2} + \cdots
&
\sigma \gg 1
\\
\frac{3}{8}(1+\frac{5}{\sigma}+\frac{35}{4\sigma^2}) + \cdots
&
\sigma \ll -1
\end{array}
\right..
\end{eqnarray}

\end{appendix}


\begin{thebibliography}{100}

\bibitem{stowoh48}
E.~C. Stoner and E.~P. Wohlfarth, Philos. Trans. Roy. Soc. London Ser. A {\bf
  240},  599  (1948).

\bibitem{nee49}
L. N{\'{e}}el, Ann. Geophys. {\bf 5},  99  (1949).

\bibitem{wernsdorfer2001}
W. Wernsdorfer, Adv. Chem. Phys. {\bf 118},  99  (2001).

\bibitem{canmyd72}
V. Cannella and J.~A. Mydosh, Phys. Rev. B {\bf 6},  4220  (1972).

\bibitem{bro63}
W.~F. Brown, Jr., Phys. Rev. {\bf 130},  1677  (1963).

\bibitem{weretal97TA}
W. Wernsdorfer, E.~B. Orozco, K. Hasselbach, A. Benoit, B. Barbara, N. Demoncy,
  A. Loiseau, H. Pascard, and D. Mailly, Phys. Rev. Lett. {\bf 78},  1791
  (1997).

\bibitem{dorfiotro97}
J.~L. Dormann, D. Fiorani, and E. Tronc, Adv. Chem. Phys. {\bf 98},  283
  (1997).

\bibitem{garpal2000acp}
J.~L. Garc{\'{\i}}a-Palacios, Adv. Chem. Phys. {\bf 112},  1  (2000).

\bibitem{batlab2002}
X. Batlle and A. Labarta, J. Phys. D: Appl. Phys. {\bf 35},  R15  (2002).

\bibitem{zalcie93}
M.~A. Za{\l}uska-Kotur and M. Cieplak, Europhys. Lett. {\bf 23},  85  (1993).

\bibitem{zal96}
M.~A. Za{\l}uska-Kotur, Phys. Rev. B {\bf 54},  1064  (1996).

\bibitem{andetal97}
J.-O. Andersson, C. Djurberg, T. Jonsson, P. Svedlindh, and P. Nordblad, Phys.
  Rev. B {\bf 56},  13\,983  (1997).

\bibitem{bergor2001}
D.~V. Berkov and N.~L. Gorn, J.~Phys.: Condens. Matter {\bf 13},  9369  (2001).

\bibitem{chikazumi}
S. Chikazumi, {\em Physics of Ferromagnetism}, 2nd ed. (Oxford Univ.
  Press, Oxford, 1997).

\bibitem{gazetal98}
F. Gazeau, J.~C. Bacri, F. Gendron, R. Perzynski, Y.~L. Ra{\u{\i}}kher, V.~I.
  Stepanov, and E. Dubois, J.~Magn. Magn. Mater. {\bf 186},  175  (1998).

\bibitem{upasrimeh2000}
R.~V. Upadhyay, D. Srinivas, and R.~V. Mehta, J.~Magn. Magn. Mater. {\bf 214},
  105  (2000).

\bibitem{bealiv59}
C.~P. Bean and J.~D. Livingston, J.~Appl. Phys. {\bf 30},  120s  (1959).

\bibitem{luoetal91}
W. Luo, S.~R. Nagel, T.~F. Rosenbaum, and R.~E. Rosensweig, Phys. Rev. Lett.
  {\bf 67},  2721  (1991).

\bibitem{garpallaz97}
J.~L. Garc{\'{\i}}a-Palacios and F.~J. L{\'{a}}zaro, Phys. Rev. B {\bf 55},
  1006  (1997).

\bibitem{raiste97}
Y.~L. Ra{\u{\i}}kher and V.~I. Stepanov, Phys. Rev. B {\bf 55},  15\,005
  (1997).

\bibitem{garjonsve2000}
J.~L. Garc{\'{\i}}a-Palacios, P. J{\"{o}}nsson, and P. Svedlindh, Phys. Rev. B
  {\bf 61},  6726  (2000).

\bibitem{hukluc2000}
B. Huke and M. L{\"u}cke, Phys. Rev. E {\bf 62},  6875  (2000).

\bibitem{jongar2001PRB}
P.~E. J{\"o}nsson and J.~L. Garc\'{\i}a-Palacios, Phys. Rev. B {\bf 64},
  174416  (2001).

\bibitem{hanjohmor98}
M. Hanson, C. Johansson, and S. M{\o}rup, Phys. Rev. Lett. {\bf 81},  735
  (1998).

\bibitem{tejetal98}
J. Tejada, X.~X. Zhang, E. del Barco, J.~M. Hern\'{a}ndez, and E.~M.
  Chudnovsky, Phys. Rev. Lett. {\bf 81},  736  (1998).

\bibitem{mamnakfur2002}
H. Mamiya, I. Nakatani, and T. Furubayashi, Phys. Rev. Lett. {\bf 88},  067202
  (2002).

\bibitem{pie33}
R. Pierls, Z. Phys. {\bf 80},  763  (1933).

\bibitem{lanlif5}
L.~D. Landau and E.~M. Lifshitz, {\em Statistical Physics}, 2nd ed. (Pergamon
  Press, Oxford, 1970).

\bibitem{wal36}
I. Waller, Z. Phys. {\bf 104},  132  (1936).

\bibitem{vanvle37}
J.~H. van Vleck, J.~Chem. Phys. {\bf 5},  320  (1937).

\bibitem{prosta98}
N.~V. Prokof'ev and P.~C.~E. Stamp, Phys. Rev. Lett. {\bf 80},  5794  (1998).

\bibitem{gri68}
R.~B. Griffiths, Phys. Rev. {\bf 176},  655  (1968).

\bibitem{banetal98}
S. Banerjee, R.~B. Griffiths, and M. Widom, J. Stat. Phys. {\bf 93},  109
  (1998).

\bibitem{pople53}
J.~A. Pople, Phil. Mag. {\bf 44},  1276  (1953).

\bibitem{nowetal2000}
U. Nowak, R.~W. Chantrell, and E.~C. Kennedy, Phys. Rev. Lett. {\bf 84},  163
  (2000).

\bibitem{cofetal98prb}
W.~T. Coffey, D.~S.~F. Crothers, J.~L. Dormann, L.~J. Geoghegan, and E.~C.
  Kennedy, Phys. Rev. B {\bf 58},  3249  (1998).

\bibitem{cofetal98jpcm}
W.~T. Coffey, D.~S.~F. Crothers, J.~L. Dormann, L.~J. Geoghegan, E.~C. Kennedy,
  and W. Wernsdorfer, J.~Phys.: Condens. Matter {\bf 10},  9093  (1998).

\bibitem{garetal99}
D.~A. Garanin, E.~C. Kennedy, D.~S.~F. Crothers, and W.~T. Coffey, Phys. Rev. E
  {\bf 60},  6499  (1999).

\bibitem{cofetal98prl}
W.~T. Coffey, D.~S.~F. Crothers, J.~L. Dormann, Y.~P. Kalmykov, E.~C. Kennedy,
  and W. Wernsdorfer, Phys. Rev. Lett. {\bf 80},  5655  (1998).

\bibitem{cofetal2001}
W.~T. Coffey, D.~S.~F. Crothers, Y.~P. Kalmykov, and S.~V. Titov, Phys. Rev. B
  {\bf 64},  012411  (2001).

\bibitem{garsve2000}
J.~L. Garc{\'{\i}}a-Palacios and P. Svedlindh, Phys. Rev. Lett. {\bf 85},  3724
   (2000).

\bibitem{dorbesfio88}
J.~L. Dormann, L. Bessais, and D. Fiorani, J.~Phys. C {\bf 21},  2015  (1988).

\bibitem{mortro94}
S. M{\o}rup and E. Tronc, Phys. Rev. Lett. {\bf 72},  3278  (1994).

\bibitem{jongar2001EPL}
P.~E. J{\"o}nsson and J.~L. Garc\'{\i}a-Palacios, Europhys. Lett. {\bf 55},
  418  (2001).

\bibitem{kubhas70}
R. Kubo and N. Hashitsume, Prog. Theor. Phys. {\bf 46},  210  (1970).

\bibitem{cofcrekal93}
W.~T. Coffey, P.~J. Cregg, and Y.~P. Kalmykov, Adv. Chem. Phys. {\bf 83},  263
  (1993).

\bibitem{gil55}
T.~L. Gilbert, Phys. Rev. {\bf 100},  1243  (1955).

\bibitem{lanlif35}
L.~D. Landau and E.~M. Lifshitz, Z. Phys. Sowjet. {\bf 8},  153  (1935).

\bibitem{raiste94PRB}
Y.~L. Ra{\u{\i}}kher and V.~I. Stepanov, Phys. Rev. B {\bf 50},  6250  (1994).

\bibitem{kub66}
R. Kubo, Rep. Prog. Phys. {\bf 29},  255  (1966).

\bibitem{garpallaz98}
J.~L. Garc{\'{\i}}a-Palacios and F.~J. L{\'{a}}zaro, Phys. Rev. B {\bf 58},
  14\,937  (1998).

\bibitem{aha69}
A. Aharoni, Phys. Rev. {\bf 177},  793  (1969).

\bibitem{cofetal2000}
W.~T. Coffey, D.~A. Garanin, H. Kachkachi, and D.~J. McCarthy, J.~Magn. Magn.
  Mater. {\bf 221},  110  (2000).

\bibitem{luietal2002}
F. Luis, F. Petroff, J.~M. Torres, L.~M. Garc{\'{\i}}a, J. Bartolom{\'e}, J.
  Carrey, and A. Vaur{\`e}s, Phys. Rev. Lett. {\bf 88},  217205  (2002).

\bibitem{hanmor98}
M.~F. Hansen and S. M{\o}rup, J.~Magn. Magn. Mater. {\bf 184},  262  (1998).

\bibitem{dorfiotro99}
J.~L. Dormann, D. Fiorani, and E. Tronc, J.~Magn. Magn. Mater. {\bf 202},  251
  (1999).

\bibitem{ewa21}
P. Ewald, Ann. Phys. {\bf 64},  253  (1921).

\bibitem{mad18}
E. Madelung, Phys. Z. {\bf 19},  524  (1918).

\bibitem{deleeetal80}
S.~W. de~Leeuw, J.~W. Perram, and E.~R. Smith, Proc. Roy. Soc. Lond. A {\bf 373},
   27  (1980).

\bibitem{jonetal95}
T. Jonsson, J. Mattsson, C. Djurberg, F.~A. Khan, P. Nordblad, and P.
  Svedlindh, Phys. Rev. Lett. {\bf 75},  4138  (1995).

\bibitem{mametal99}
H. Mamiya, I. Nakatani, and T. Furubayashi, Phys. Rev. Lett. {\bf 82},  4332
  (1999).

\bibitem{doretal99}
J.~L. Dormann, D. Fiorani, R. Cherkaoui, E. Tronc, F. Lucari, F. D'Orazio, L.
  Spinu, M. Nogu{\`e}s, H. Kachkachi, and J.~P. Jolivet, J.~Magn. Magn. Mater.
  {\bf 203},  23  (1999).

\bibitem{djuetal97}
C. Djurberg, P. Svedlindh, P. Nordblad, M.~F. Hansen, F. B{\o}dker, and S.
  M{\o}rup, Phys. Rev. Lett. {\bf 79},  5154  (1997).

\bibitem{jonsvehan98}
T. Jonsson, P. Svedlindh, and M.~F. Hansen, Phys. Rev. Lett. {\bf 81},  3976
  (1998).

\bibitem{hanetal2002}
M.~F. Hansen, P. J{\"o}nsson, P. Nordblad, and P. Svedlindh, J.~Phys.: Condens.
  Matter {\bf 14},  4901  (2002).

\bibitem{sahetal2002}
S. Sahoo, O. Petracic, C. Binek, W. Kleemann, J.~B. Sousa, S. Cardoso, and
  P.~P. Freitas, Phys. Rev. B {\bf 65},  134406  (2002).

\bibitem{binyou86}
K. Binder and A.~P. Young, Rev. Mod. Phys. {\bf 58},  801  (1986).

\bibitem{fischerhertz}
K.~H. Fischer and J.~A. Hertz, {\em Spin Glasses} (Cambridge Univ. Press,
  Cambridge, U.K., 1991).

\bibitem{young}
 A.~P. Young, ed., {\em Spin Glasses and Random Fields}, (World Scientific,
  Singapore, 1997).

\bibitem{rudkit54}
M.~A. Ruderman and C. Kittel, Phys. Rev. {\bf 96},  99  (1954).

\bibitem{kas56}
T. Kasuya, Prog. Theor. Phys. {\bf 16},  45 and 58  (1956).

\bibitem{yos57}
K. Yosida, Phys. Rev. {\bf 106},  893  (1957).

\bibitem{ferlev80}
A. Fert and P.~M. Levy, Phys. Rev. Lett. {\bf 44},  1538  (1980).

\bibitem{prejol80}
J.~J. Prejean, M.~J. Joliclerc, and P. Monod, J. Physique {\bf 41},  427
  (1980).

\bibitem{itoetal86}
A. Ito, H. Aruga, E. Torikai, M. Kikuchi, Y. Syono, and H. Takei, Phys. Rev.
  Lett. {\bf 57},  483  (1986).

\bibitem{itoetal90}
A. Ito, E. Torikai, S. Morimoto, H. Aruga, M. Kikuchi, Y. Syono, and H. Takei,
  J.~Phys. Soc. Jpn. {\bf 59},  829  (1990).

\bibitem{aruito93}
H. {Aruga Katori} and A. Ito, J.~Phys. Soc. Jpn. {\bf 62},  4488  (1993).

\bibitem{edwand75}
S.~F. Edwards and P.~W. Anderson, J.~Phys. F {\bf 5},  965  (1975).

\bibitem{shekir75}
D. Sherrington and S. Kirkpatrick, Phys. Rev. Lett. {\bf 35},  1792  (1975).

\bibitem{par79}
G. Parisi, Phys. Rev. Lett. {\bf 43},  1754  (1979).

\bibitem{par80}
G. Parisi, J.~Phys. A {\bf 13},  1101  (1980).

\bibitem{par83}
G. Parisi, Phys. Rev. Lett. {\bf 50},  1946  (1983).

\bibitem{suzuki77}
M. Suzuki, Prog. Theor. Phys. {\bf 58},  1151  (1977).

\bibitem{chietal79}
S. Chikazawa, T. Saito, T. Sato, and Y. Miyako, J.~Phys. Soc. Jpn. {\bf 47},
  335  (1979).

\bibitem{norsve97}
P. Nordblad and P. Svedlindh,  in {\em Spin Glasses and Random Fields}, A.~P. Young, ed. (World Scientific, Singapore, 1997), pp.\ 1--27.

\bibitem{petit-thesis}
D. Petit, Ph.D. thesis, Universit{\'e} Paris XI, 2002.

\bibitem{jonetal2002PRL}
P.~E. J{\"o}nsson, H. Yoshino, P. Nordblad, H.~A. Katori, and A. Ito, Phys.
  Rev. Lett. {\bf 88},  257204  (2002).

\bibitem{struik78}
L.~C.~A. Struik, {\em Physical Aging in Amorphous Polymers and Other Materials}
  (Elsevier, Amsterdam, 1978).

\bibitem{lunetal83}
L. Lundgren, P. Svedlindh, P. Nordblad, and O. Beckman, Phys. Rev. Lett. {\bf
  51},  911  (1983).

\bibitem{andmatsve92}
J.~O. Andersson, J. Mattsson, and P. Svedlindh, Phys. Rev. B {\bf 46},  8297
  (1992).

\bibitem{belcillar2000}
L. Bellon, S. Ciliberto, and C. Laroche, Europhys. Lett. {\bf 51},  551
  (2000).

\bibitem{douetal99}
P. Doussineau, T. de~Lacerda-Ar{\^o}so, and A. Levelut, Europhys. Lett. {\bf
  46},  401  (1999).

\bibitem{noretal2000}
V. Normand, S. Muller, J.-C. Ravey, and A. Parker, Macromolecules {\bf 33},
  1063  (2000).

\bibitem{papetal99}
E.~L. Papadopoulou, P. Nordblad, P. Svedlindh, R. Sch{\"o}neberger, and R.
  Gross, Phys. Rev. Lett. {\bf 82},  173  (1999).

\bibitem{magetal97}
J. Magnusson, C. Djurberg, P. Granberg, and P. Nordblad, Rev. Sci. Instrum.
  {\bf 68},  3761  (1997).

\bibitem{mcm84}
W.~L. McMillan, J.~Phys. C {\bf 17},  3179  (1984).

\bibitem{bramor86}
A.~J. Bray and M.~A. Moore,  in {\em Heidelberg Colloquium on Glassy Dynamics},
 J.~L. van Hemmen and I. Morgenstern, eds. (Springer, Berlin, 1986).

\bibitem{fishus86}
D.~S. Fisher and D.~A. Huse, Phys. Rev. Lett. {\bf 56},  1601  (1986).

\bibitem{fishus88eq}
D.~S. Fisher and D.~A. Huse, Phys. Rev. B {\bf 38},  386  (1988).

\bibitem{fishus88noneq}
D.~S. Fisher and D.~A. Huse, Phys. Rev. B {\bf 38},  373  (1988).

\bibitem{bouetal2001}
J.-P. Bouchaud, V. Dupuis, J. Hammann, and E. Vincent, Phys. Rev. B {\bf 65},
  024439  (2001).

\bibitem{dupetal2001}
V. Dupuis, E. Vincent, J.-P. Bouchaud, J. Hammann, A. Ito, and H. {Aruga
  Katori}, Phys. Rev. B {\bf 64},  174204  (2001).

\bibitem{jonyosnor2002}
P. J{\"o}nsson, H. Yoshino, and P. Nordblad, Phys. Rev. Lett. {\bf 89},  097201
   (2002).

\bibitem{beretal}
L. Berthier, V. Viasnoff, O. White, V. Orlyanchik, and F. Krzakala, cond-mat/0211106.

\bibitem{bramoo84}
A.~J. Bray and M.~A. Moore, J.~Phys. C {\bf 17},  L463  (1984).

\bibitem{risken}
H. Risken, {\em The Fokker--Planck Equation}, 2nd ed. (Springer, Berlin, 1989).

\bibitem{yoshuktak2002}
H. Yoshino, K. Hukushima, and H. Takayama, Phys. Rev. B {\bf 66},  064431
  (2002).

\bibitem{bramoo87}
A.~J. Bray and M.~A. Moore, Phys. Rev. Lett. {\bf 58},  57  (1987).

\bibitem{graetal88}
P. Granberg, L. Sandlund, P. Nordblad, P. Svedlindh, and L. Lundgren, Phys.
  Rev. B {\bf 38},  7079  (1988).

\bibitem{jonetal98}
K. Jonason, E. Vincent, J. Hammann, J.-P. Bouchaud, and P. Nordblad, Phys. Rev.
  Lett. {\bf 81},  3243  (1998).

\bibitem{jonetal99}
T. Jonsson, K. Jonason, P. J{\"o}nsson, and P. Nordblad, Phys. Rev. B {\bf 59},
   8770  (1999).

\bibitem{matetal2001}
R. Mathieu, P. J{\"o}nsson, D.~N.~H. Nam, and P. Nordblad, Phys. Rev. B {\bf
  63},  092401  (2001).

\bibitem{matetal2002}
R. Mathieu, P.~E. J{\"o}nsson, P. Nordblad, H.~A. Katori, and A. Ito, Phys.
  Rev. B {\bf 65},  012411  (2002).

\bibitem{jonhannor2000}
P. J{\"o}nsson, M.~F. Hansen, and P. Nordblad, Phys. Rev. B {\bf 61},  1261
  (2000).

\bibitem{coletal2000}
E.~V. Colla, L.~K. Chao, M.~B. Weissman, and D.~D. Viehland, Phys. Rev. Lett.
  {\bf 85},  3033  (2000).

\bibitem{belcillar2002}
L. Bellon, S. Ciliberto, and C. Laroche, Eur. Phys. J.~B {\bf 25},  223
  (2002).

\bibitem{kitetal2002}
A.~V. Kityk, M.~C. Rheinst{\"a}dter, K. Knorr, and H. Rieger, Phys. Rev. B {\bf
  65},  144415  (2002).

\bibitem{garetal2003}
A. Gardchareon, R. Mathieu, P.~E. J{\"o}nsson, and P. Nordblad, Phys. Rev. B
  {\bf 67},  052505  (2003).

\bibitem{jonetal2003}
P.~E. J{\"o}nsson, R. Mathieu, H. Yoshino, P. Nordblad, H.~A. Katori, and A.
  Ito, cond-mat/0307640.

\bibitem{jonetal2001JMMM}
P. J{\"o}nsson, P. Svedlindh, P. Nordblad, and M.~F. Hansen, J.~Magn. Magn.
  Mater. {\bf 226-230},  1315  (2001).

\bibitem{jonetal2001PRB}
P. J{\"o}nsson, S.Felton, P. Svedlindh, P. Nordblad, and M.~F. Hansen, Phys.
  Rev. B {\bf 64},  212402  (2001).

\bibitem{hanetal95a}
M. Hanson, C. Johansson, and S. M{\o}rup, J. of Phys.: Condens. Matter
  {\bf 7},  9263  (1995).

\bibitem{hanetal95b}
M. Hanson, C. Johansson, M.~S. Pedersen, and S. M{\o}rup, J. of Phys.:
  Condens. Matter {\bf 7},  9269  (1995).

\bibitem{hanetal98}
M.~F. Hansen, F. B{\o}dker, S. M{\o}rup, C. Djurberg, and P. Svedlindh,
  J.~Magn. Magn. Mater. {\bf 177-181},  928  (1998).

\bibitem{komyostak2000A}
T. Komori, H. Yoshino, and H. Takayama, J.~Phys. Soc. Jpn. {\bf 69 {\rm Suppl.
  A}},  228  (2000).

\bibitem{picricrit2001}
M. Picco, F. Ricci-Tersenghi, and F. Ritort, Phys. Rev. B {\bf 63},  174412
  (2001).

\bibitem{berbou2002}
L. Berthier and J.~P. Bouchaud, Phys. Rev. B {\bf 66},  054404  (2002).

\bibitem{matetal93}
J. Mattsson, C. Djurberg, P. Nordblad, L. Hoines, R. Stubi, and J.~A. Cowen,
  Phys. Rev. B {\bf 47},  14626  (1993).

\bibitem{schetal93prb}
A.~G. Schins, E.~M. Dons, A.~F.~M. Arts, H.~W. Wijn, E. Vincent, L. Leylekian,
  and J. Hammann, Phys. Rev. B {\bf 48},  16524  (1993).

\bibitem{rigaux95}
C. Rigaux, Ann. Phys. Fr. {\bf 20},  445  (1995).

\bibitem{ogielski85}
A.~T. Ogielski, Phys. Rev. B {\bf 32},  7384  (1985).

\bibitem{gunetal88}
K. Gunnarsson, P. Svedlindh, P. Nordblad, L. Lundgren, H. Aruga, and A. Ito,
  Phys. Rev. Lett. {\bf 61},  754  (1988).

\bibitem{jonnorsve98}
T. Jonsson, P. Nordblad, and P. Svedlindh, Phys. Rev. B {\bf 57},  497  (1998).

\bibitem{ros-lax52}
R. Rosenberg and M. Lax, J.~Chem. Phys. {\bf 21},  424  (1952).

\bibitem{mathews-walker}
J. Mathews and R.~L. Walker, {\em Mathematical Methods of Physics}, 2nd ed.
  (Benjamin/Cummings, Menlo Park, California, 1970).

\bibitem{recipes}
W.~H. Press, S.~A. Teukolsky, W.~T. Vetterling, and B.~P. Flannery, {\em
  Numerical Recipes}, 2nd ed. (Cambridge Univ. Press, New York, 1992).

\bibitem{kalcof97}
Y.~P. Kalmykov and W.~T. Coffey, Phys. Rev. B {\bf 56},  3325  (1997).

\end{thebibliography}
\end{document}